\documentclass{article}
\usepackage{arxiv}
\usepackage[utf8]{inputenc} 
\usepackage[T1]{fontenc}    
\usepackage{hyperref}      
\usepackage{url}            
\usepackage{booktabs}       
\usepackage{amsfonts}       
\usepackage{nicefrac}       
\usepackage{microtype}      
\usepackage{mathtools}         
\usepackage{graphicx}
\usepackage{natbib}
\usepackage{doi}
\usepackage{bm}
\usepackage{bbm}
\usepackage{amsmath,mathrsfs,amssymb, array, dsfont}
\usepackage{cleveref}
\usepackage{empheq}
\usepackage{enumerate}
\usepackage[shortlabels]{enumitem}
\usepackage{caption}

\makeatletter
\renewenvironment{cases}[1][l]{\matrix@check\cases\env@cases{#1}}{\endarray\right.}
\def\env@cases#1{%
  \let\@ifnextchar\new@ifnextchar
  \left\lbrace\def\arraystretch{1.2}%
  \array{@{}#1@{\quad}l@{}}}
\makeatother

\makeatletter
\newcases{dcasesnoquad}{}{%
  $\m@th\displaystyle{##}$\hfil}
  {$\m@th\displaystyle{{}##}$\hfil}{\lbrace}{.}
\makeatother

\newcommand{\beginsupplement}{%
        \setcounter{table}{0}
        \renewcommand{\thetable}{S\arabic{table}}%
        \setcounter{figure}{0}
        \renewcommand{\thefigure}{S\arabic{figure}}%
     }
\def\myformat#1{\centering#1}

\newtheorem{theorem}{Theorem}
\newtheorem{lemma}[theorem]{Lemma}

\title{Bayesian Age Category Reconciliation for Age- and Cause-specific Under-five Mortality Estimates}

\date{}

\author{\hspace{1mm}Shuxian Fan \\
	University of Washington, USA\\
	\texttt{fansx@uw.edu} \\
	\AND
	\hspace{1mm}Li Liu \\
	Johns Hopkins University, USA\\
	\texttt{liliu@jhsph.edu} \\
	\AND
	\hspace{1mm}Jamie Perin\\
	Johns Hopkins University, USA\\
	\texttt{jperin@jhu.edu}\\
	\AND
	\hspace{1mm} Tyler H. McCormick\\
	University of Washington, USA\\
	\texttt{tylermc@uw.edu}
}

\renewcommand{\headeright}{}
\renewcommand{\undertitle}{}
\renewcommand{\shorttitle}{}

\hypersetup{
pdftitle={Bayesian Age Category Reconciliation for Age- and Cause-specific Under-five Mortality Estimates},
pdfsubject={q-bio.NC, q-bio.QM},
pdfauthor={Shuxian Fan, Li Liu, Jamie Perin, Tyler H. McCormick},
}

\begin{document}
\maketitle

\begin{abstract}
Age-disaggregated health data is crucial for effective public health planning and monitoring.  Monitoring under-five mortality, for example, requires highly detailed age data since the distribution of potential causes of death varies substantially within the first few years of life. Comparative researchers often have to rely on multiple data sources yet, these sources often have ages aggregated at different levels, making it difficult to combine the data into a single, coherent picture. To address this challenge in the context of under-five cause-specific mortality, we propose a Bayesian approach, that calibrates data with different age structures to produce unified and accurate estimates of the standardized age group distributions. We consider age-disaggregated death counts as fully-classified multinomial data and show that by incorporating partially-classified aggregated data, we can construct an improved Bayes estimator of the multinomial parameters under the Kullback-Leibler (KL) loss. We illustrate the method using both synthetic and real data, demonstrating that the proposed method achieves adequate performance in imputing incomplete classification. Finally, we present the results of numerical studies examining the conditions necessary for obtaining improved estimators. These studies provide insights and interpretations that can be used to aid future research and inform guidance for practitioners on appropriate levels of age disaggregation, with the aim of improving the accuracy and reliability of under-five cause-specific mortality estimates.
\end{abstract}

\section{Introduction}
\label{intro}

Age- and cause-specific under-five mortality (ACSU5M) rates are a critical indicator of the health and well-being of children worldwide. Therefore, understanding patterns and trends in ACSU5M is crucial for informing and evaluating age and disease-targeted interventions and policies aimed at reducing child mortality. 

ACSU5M data are typically collected through demographic and health surveys, sample registration systems (SRS), and vital registration (VR) systems. A persistent challenge for accurate and reliable ACSU5M estimation is the lack of complete and consistent data, particularly in low-income countries, where VRs are often incomplete or unavailable. In the absence of individual-level registration data, researchers often have to rely on multiple data sources to estimate ACSU5M, such as SRSs, household surveys, and verbal autopsy (VA) questionnaires. These data are often provided in an aggregated manner, with ages grouped into various levels, making it difficult to perform age-sensitive analysis as it requires standardized age-disaggregated death counts~\citep{diaz2021call}. Despite the importance of data reconciliation in child mortality research, particularly in ACSU5M studies, the theoretical support for the appropriate data disaggregation approach has not been fully developed. In many cases, researchers face a shortage of individual death data with exact recorded ages and must rely on aggregated data where ages are grouped. When dealing with aggregated data, researchers often resort to past studies in choosing the appropriate age categories to use, without considering the empirical evidence. This can result in the use of age groupings that may not accurately reflect the true age distribution within each cause of death (COD). Table~\ref{tab:age} gives an example of typical breakdowns in age categories at increasing levels of disaggregation.

\begin{table}[!htb]
\centering
\includegraphics[width=0.5\linewidth]{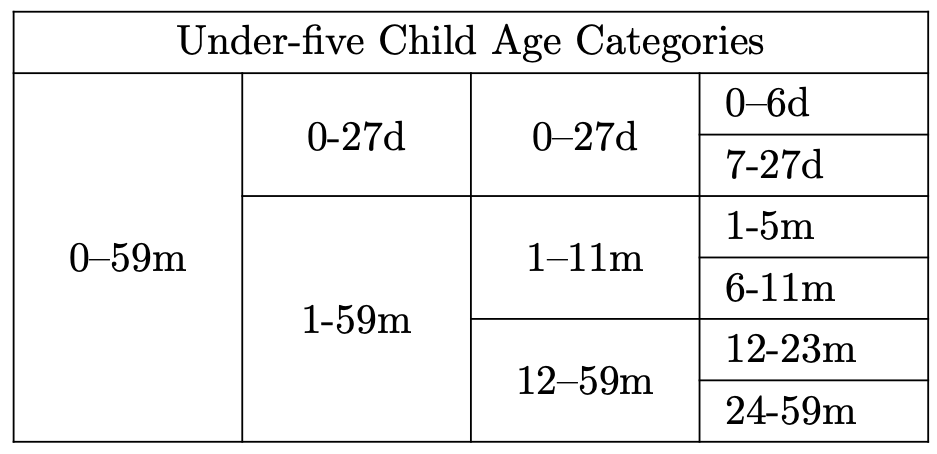}
\vspace{2mm}
\caption{Age category disaggregation in ACSU5M studies.}
\label{tab:age}
\end{table}

This paper proposes a Bayesian approach to calibrate across data sources reported at different levels of disaggregation and provides estimates of standardized age group distributions. The method combines both individual registration data, if available, and fully-classified age-disaggregated death counts as group counts from a multinomial distribution. The partially-classified aggregated data are then incorporated to jointly estimate the multinomial probabilities, potentially resulting in improved estimates of the age group distributions as well as age- and cause-specific death counts at the desired level. 

The problem of concurrently estimating multinomial probabilities has been widely studied in the literature, with numerous Bayesian methods proposed~\citep{fienberg1973simultaneous,leonard1977bayesian,alam1986empirical,albert1987empirical}. The case of partial classification in contingency tables has also been explored~\citep{chen1976analysis,albert1985bayesian,gibbons1994bayesian}, where data are partially classified by rows or columns.~\cite{ahn2010bayesian} expanded upon these ideas by proposing a Bayesian approach to handle incompletely classified multinomial data in the study of pathogen diversity. However, the decision-theoretic foundations of these methods have not been fully established. Unlike the previous approaches that limit the partial classification to row (column) sums, our method considers a more general scenario, where partial classifications can be any partitions of the fully-classified group index set and provides conditions that incorporating partially-classified data can lead to better Bayes estimators. Additionally, we tailor the method to address the age category reconciliation problem in ACSU5M studies in a more general setting,  considering scenarios where the age groups are not completely nested between data sources, as opposed to the nested age structures depicted in Table~\ref{tab:age}. In both cases, we provide comprehensive frameworks for conducting Bayesian inference. Our contributions are as follows:
\begin{itemize}
    \item Our work extends the existing literature on the simultaneous estimation of multinomial probabilities to a more versatile setting. This allows for greater applicability to a wider range of problems. 
    \item We provide theoretical support and numerical studies from the decision-theoretic perspective, demonstrating how the integration of partially classified data can result in enhanced Bayesian estimators of multinomial probabilities under certain conditions, which we explicitly define. 
    \item We conduct simulation studies based on observed, disaggregated data to assess the effectiveness of our age reconciliation method. Our results demonstrate that the proposed approach is promising, and offers novel and valuable perspectives on the mitigation of age inconsistencies in ACSU5M studies. 
\end{itemize}

\section{Method}

\subsection{Problem Statement}

Let $\bm X = (X_i)_{i=1}^k$ be the fully classified observations that follow a multinomial distribution with parameters $(N, \bm \theta)$,

\[\bm X \sim \text{Multi}(\bm x | N, \bm\theta)  = \frac{N!}{\prod_{i=1}^k x_i!} \prod_{i=1}^k \theta_i^{x_i},\]
where $\bm \theta = \{\theta_i\}_{i=1}^k$ is the unknown vector of group probabilities and $\bm x = (x_i)_{i=1}^k\in \{(x'_i)_{i=1}^k \in \mathbb{N}_0|\sum_{i=1}^k x'_i = N\}$,  $\mathbb{N}_0 = \{0\} \cup \mathbb{N} = \{0, 1, 2, \dots\}$. In the ACU5M case, for example, this could be the age- and cause-spefic death counts we observe at a desirable disaggregated level with $k$ different groups. Suppose that we have additional data that are partially classified with respect to age levels, that is, $\bm Y'= (Y'_i)_{i=1}^k\sim \mbox{Multi}(N', \bm \theta)$ independently of $\bm X$, but we only observe the partially classified data 
$\bm Y = (Y_{A_j})_{j=0}^m$ that follow a multinomial distribution with parameters $(N', \bm \tau)$,
where $(A_j)_{j=1}^m$ are the distinct non-singleton proper subsets of $S = \{1,\dots,k\}$, $A_0 = S - \cup_{j=1}^m A_j$, and $\bm \tau = (\tau_j)_{j=0}^m$ with $\tau_j = \sum_{i\in A_j}\theta_i$. Note that $A_0$ can be $\emptyset$. An example of an age disaggregation setting using the notation introduced above is presented below:
 \begin{equation*}
\begin{dcasesnoquad}
\mbox{\{0-27d\}}:= \{1, 2\} &\begin{cases}[r]\mbox{\{0-6d\}}:=\{1\} &\\ \mbox{\{7-27d\}}:=\{ 2 \}&\end{cases}\\
\mbox{\{1-11m\}}:= \{3, 4\} & \begin{cases}[r] \mbox{\{1-5m\}} :=\{3\}& \\
      \mbox{\{6-11m\}}:=\{ 4 \}&\end{cases}\\
\mbox{\{12-59m\}}:= \{5, 6\}&\begin{cases} [r]\mbox{\{12-23m\}}:=\{ 5 \} & \\
          \mbox{\{24-59m\}} :=\{ 6 \}&\end{cases},
\end{dcasesnoquad}
\end{equation*}
from which we have that
\begin{align*}
S &= \{1, 2, 3, 4, 5, 6\},\\
A_1 & = \{1,2\}, A_2 = \{3, 4\}, A_3 = \{5, 6\}.
\end{align*}

We consider the problem of estimating $\bm \theta$ with the prior distribution for $\bm \theta$ being the Dirichlet distribution with parameter $\bm \alpha = (\alpha_i)_{i=1}^k$ under the KL loss that can be directly interpreted as divergence measures induced by entropy~\citep{nayak1989estimating},
\[L(\bm p, \bm \theta) = \sum_{i=1}^k \theta_i \log \frac{\theta_i}{p_i},\]
where $\bm p = (p_i)_{i=1}^k \in \Theta =  \{(\theta_i)_{i=1}^k\in (0,\infty)^{k}|\sum_{i=1}^k \theta_i = 1\}.$ The problem is further broken down into two scenarios: (i) where the $A_j$ sets are mutually exclusive; and (ii) where the $A_j$ sets may overlap. 

\subsection{Bayes Estimator with Disjoint Partial Classifications}

Let $\bm\rho_{A_j} = (\theta_i/\tau_j,i\in A_j)^T$, $j = 0, \dots, m$ be the conditional probabilities of each individual cell within each group $A_j$ and $\bm \eta = (\bm \tau, \bm \rho_{A_j}, j = 0,\dots, m)$. Denote $(\bm X, \bm Y) = (X_1,X_2,\dots, X_k, Y_{A_1}, \dots, Y_{Am})$ and the probability mass function of $(\bm X, \bm Y) = (\bm x,\bm y)$ is given by
\begin{align*}
& f(\bm x, \bm y|\bm\eta) = \\
& \frac{(N!)(N'!)}{(\prod_{i=1}^k x_i!)(\prod_{j=0}^m y_{A_j}!)}\prod_{j=0}^m \tau_j^{x_{A_j} + y_{A_j}}\prod_{j = 0}^m\prod_{r\in A_j}(\rho_{A_j}^{(r)})^{x_r},
\end{align*}
where $x_{A_j} = \sum_{r\in A_j} x_r$. 

Consider the Bayes estimator with the prior distribution for $\bm \theta$ being the Dirichlet distribution with parameter $\bm \alpha$, which is
\begin{align*}
    f(\bm\theta|\bm\alpha) \propto \prod_{i=1}^k \theta_i^{\alpha_i - 1}.
\end{align*}
We can show that the posterior density function is given by
\begin{equation}
\begin{aligned}
f(\bm \eta|\bm x, \bm y) &\propto \tau_{0}^{x_{A_0}+\alpha_{A_0}-1}\prod_{j=1}^m \tau_{j}^{x_{A_j} + y_{A_j}+\alpha_{A_j} -1}\\
&\times \prod_{j=0}^m\prod_{r\in A_j}\left(\rho_{A_j}^{(r)}\right)^{x_r + \alpha_r - 1},
\end{aligned}
\label{eq:1}
\end{equation}
where $\alpha_{A_j} = \sum_{r\in A_j}\alpha_r$.

Since $\bm \theta$ and $\bm\eta$ bear a one-to-one relationship, they are equivalent parameterizations. It follows from~\ref{eq:1} that $\bm \tau$ and $\bm (\bm\rho_{A_j})_{j=0}^m$ are jointly independent and $(\rho_{A_j}^{(r)})_{r\in A_j = \{j_1,\dots, j_{n_j}\}}$ follows a Dirichlet distribution with parameter $(\alpha_{j_1} + x_{j_1},\dots, \alpha_{j_{n_j}} + x_{j_{n_j}})$ for $j = 0,\dots, m$. Moreover, it can be readily shown that $\bm \tau$ also follows a Dirichlet distribution with parameter $(x_{A_0} + \alpha_{A_0}, x_{A_1} + y_{A_1} + \alpha_{A_1}, \dots, x_{A_j} + y_{A_j} + \alpha_{A_j})$. The closed-form Bayes estimator under the KL loss, which is the posterior mean of $\bm \theta$ based on $(\bm x, \bm y)$ is given in Lemma~\ref{lemma1}. 
\begin{lemma}
With respect to the Dirichlet prior with parameter $\bm \alpha = (\alpha_1, \dots, \alpha_k)$, given data consist of fully-classified $\bm x = (x_i)_{i=1}^{k}$ and partially-classified $\bm y = (y_{A_j})_{j=0}^m$ with disjoint $A_j$'s, the Bayes estimator $(\hat\theta_i)_{i=1}^k$, for $i\in A_j$, is given by
\begin{align*}
\hat{\theta}_i = \frac{\alpha_i + x_i}{x_{A_j} + \alpha_{A_j}}  \frac{y_{A_j} + \alpha_{A_j} + x_{A_j}}{N + N' + \alpha_S},
\end{align*}
\label{lemma1}
\end{lemma}
where $\alpha_S = \sum_{i=1}^k \alpha_i$.

\subsection{Decision-theoretic Justification for Age Reconciliation}
In order to provide a decision-theoretic rationale for the impact of incorporating partially-classified data on the estimations of the multinomial probabilities, we compare the risk functions of $\bm \hat \theta$ and the Bayes estimator $\Tilde{\bm\theta}$ based only on the fully-classified data $\bm x$, which is given by 
\begin{align*}
    \tilde{\theta_i} = \frac{x_i + \alpha_i}{\alpha_S + N},
\end{align*}
owing to the conjugation of the Dirichlet prior for the multinomial distribution. 

Let the risk difference be $\Delta_{\bm\theta}(N, N') = \mathbb{E}_{\bm\theta}[L(\hat{\bm \theta}, \bm\theta)] - \mathbb{E}_{\bm\theta}[(L(\tilde{\bm \theta}, \bm\theta)]$. The following lemma gives the decomposition of $\Delta_{\bm\theta}(N, N')$.

\begin{lemma}
The risk difference between estimators $\hat{\bm\theta}$ and $\Tilde{\bm\theta}$ can be decomposed as
\begin{align*}
 \Delta_{\bm\theta}(N, N')  &= \mathbb{E}_{\bm \theta}\bigg[\log \frac{N' + N + \alpha_S}{N +\alpha_S }  \\
 & +\sum_{j=0}^m \theta_{A_j}\log\left\{\frac{x_{A_j} + \alpha_{A_j}}{y_{A_j} +x_{A_j} + \alpha_{A_j} }\right\}\bigg]\\
&= \sum_{u=1}^{N'}\Delta_{\bm \theta}(N + u - 1, 1) .
\end{align*}
    \label{lemma2}
\end{lemma}

All proofs can be found in the supplementary materials. As shown in Lemma~\ref{lemma1},  $\Delta_{\bm\theta}(N, N')$ can be expressed as the sum of $\Delta_{\bm \theta}(N + u - 1, 1)$, where $u = 1,\dots, N'$ represents the number of additional partially-classified data points we considered. Without loss of generality, we consider $\Delta_{\bm \theta}(N, 1)$. The following lemma gives the condition when the maximum of $\Delta_{\bm \theta}(N, 1)$ can be achieved at $\bm \theta^* = (\theta_{i}^*)_{i=0}^k \in \bm\Theta$ where for $i\in A_j = \{ j_1, \dots, j_{n_j}\}$, $\theta_{i: i\in A_j}^* = 1/[(1+m)n_{j}]$ and $\sum_{i=0}^k \theta^*_{i:i\in A_j} = \theta_{A_j} = 1/(m+1)$ for all $j = 0, 1,\dots, m$.

\begin{lemma}
 Suppose that $\min_{j}\alpha_{A_j} \geq 2$, then the risk difference $\Delta_{\bm\theta}(N,1)$ is maximized at $\bm\theta = \bm \theta^*$:
    \[\max_{\bm\theta \in \Theta} \Delta_{\bm \theta}(N, 1) = \Delta_{\bm\theta^*}(N,1).\]
    \label{lemma3}
\end{lemma}
We then establish the dominance of the estimators $\hat{\bm\theta}$ and $\tilde{\bm\theta}$ in the following theorem.

\begin{theorem}

\begin{enumerate}[(i)]
    \item \label{t1} Fix $m\in \mathbb{N}$ and $N\in \mathbb{N}$. Suppose we have $A_j$'s and $\alpha_i$'s such that $\min_{j}\alpha_{A_j} \geq 2$, then $\hat{\bm\theta}$ dominates $\Tilde{\bm\theta}$ when $N'$ is sufficiently large.
    \item \label{t2} Fix $m\in \mathbb{N}$ and $N'\in \mathbb{N}$. Suppose we have $A_j$'s and $\alpha_i$'s such that $\min_{j}\alpha_{A_j} \geq 2$, then $\hat{\bm\theta}$ dominates $\Tilde{\bm\theta}$ when $N$ is sufficiently large.
\end{enumerate}

\label{theorem1}
\end{theorem}

Theorem~\ref{theorem1} states the sufficient condition for $\Tilde{\bm \theta}$ to be dominated by $\hat{\bm \theta}$, which requires $\alpha_{A_j} = \sum_{i\in A_j}\alpha_i \geq 2$ for all $j = 0,\dots, m$. For example, if non-informative priors are used, the condition is $\min_j |A_j| \geq 2$ for the uniform prior and $\min_j |A_j| \geq 4$ for Jeffrey's prior. With $m$ being fixed, the conditions also require the number of classes $k$ to be at minimum $2(m+1)$ with the uniform prior and $4(m+1)$ with Jeffrey's prior since $\sum_{j=0}^m |A_j| = k$. Although this may seem counter-intuitive, it can be interpretable in the context of ACSU5M age reconciliation. The condition for $k$ suggests that the granularity of the age groups must be sufficient to overcome the uncertainties introduced by the partial classifications. This requirement is consistent with the nature of the ACSU5M age reconciliation problem, where the age range of the observations is fixed to be between 0 to 5 years old, and the partially-classified age groups are typically reported in an aggregated manner with a fixed $m$. Additionally, it is also worth mentioning that the value of $k$ can be regarded as a truncation level, which is closely related to the truncation bounds for the Dirichlet Process~\citep{ishwaran2001gibbs} and parallel work for the Indian buffet process~\citep{doshi2009variational}, that the bound decreases with the truncation level.

Moreover, the Dirichlet parameter vector $\bm \alpha$ captures the prior belief about $\bm \theta$. It can be seen as a pseudo-count of observations of each class before the actual data is collected~\citep{teh2010dirichlet}. The resulting Dirichlet-multinomial distribution approximates the multinomial distribution arbitrarily well for large $\bm\alpha$ values, which corresponds to strong prior knowledge about the distribution whereas small $\bm\alpha$ values correspond to weak or none prior information. This provides us with a natural framework for incorporating pre-existing knowledge about the age distributions, such as from census data, especially when individual registration or fully-classified data is limited. 

\subsection{Gibbs Sampling Scheme with Overlapping Partial Classifications}
In the context of ACSU5M studies, it is possible for data sources to exhibit non-nested age structures. This means that the age groups $A_j$ may overlap with each other, making the posterior distribution intractable. However, it is still possible to make inferences by utilizing the Gibbs sampling scheme~\citep{gelfand2000gibbs}. 

Let $((Z_{i|A_j})_{i=1}^k)_{j=0}^k$ denote the count of the partially-classified observations that belong to category $i$, and are included when counting for $A_j$. Then $Y_{A_j}$ can be written as \[Y_{A_j} = \sum_{i=1}^k Z_{i|A_j}.\]
Then the Gibbs sampling scheme can be implemented as follows, with the $t$-th iterations obtained by:
\begin{align*}
(Z_{i|A_j})_{i=1}^k  & \sim \mbox{Multi}(y_{A_j}, \bm \pi_j ) \\
c_{i}^{(t)} &= x_i + \sum_{j=0}^m z_{i|A_j}\\
\bm\theta^{(t+1)} & \sim \mbox{Dir}(\bm \alpha_t)
\end{align*}
where $\bm \pi_j = (\theta_i\mathbbm{1}\{i\in A_j\}/\theta_{A_j})_{i=1}^k$, and $\bm \alpha_t = (\alpha_i + c_i^{(t)})_{i=1}^k$.

\section{Experiments}

\subsection{Numerical Study}
In this section, we present the results of the numerical studies we conducted to investigate the finite-sample performance of $\hat{\bm \theta}$ and $\Tilde{\bm \theta}$ in various classification settings:
\begin{enumerate}[(i)]
\item \label{s1} $S = \{1, 2, 3\}$, $A_0 = \{1\}$, $A_1 = \{2, 3\}$ 
\item \label{s2} $S = \{1, 2,\dots, 9\}$, $A_0 = \{1,2,3,4\}$, $A_1 = \{5$,$6,7,8,9\}$
\item \label{s3} $S = \{1, 2, 3\}$, $A_0 = \{1,2\}$, $A_1 = \{2, 3\}$ 
\item \label{s4} $S = \{1, 2,\dots, 9\}$, $A_0 = \{1,2,3,4,5\}$, $A_1 = \{5$,$6,7,8,9\}$
\end{enumerate}

For each setting, we use Jeffrey's prior for $\bm \theta_1 = (1/3, 1/3, 1/3)$ and $\bm \theta_2 = (1/9, \dots, 1/9)$. The following sample sizes are considered:
\begin{itemize}
    \item Fix $N \in \{50, 100, 200\}$ and vary $N'\in \{50, 100$, $200, 500, 1000, 2000\}$.
    \item  Fix $N' \in \{50, 100, 200\}$ and vary $N\in \{50, 100$, $200, 500, 1000, 2000\}$
\end{itemize}

\begin{figure}[!htb]
\centering
\includegraphics[width=0.9\linewidth]{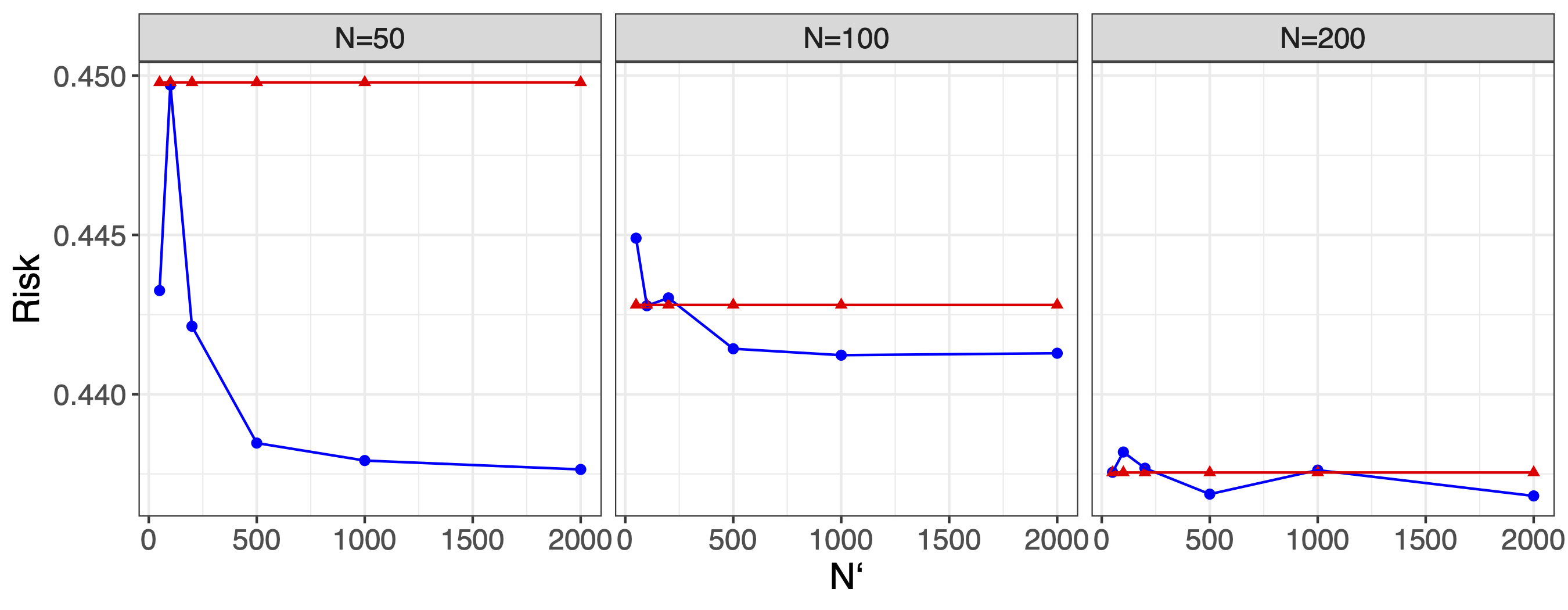}
\includegraphics[width=0.9\linewidth]{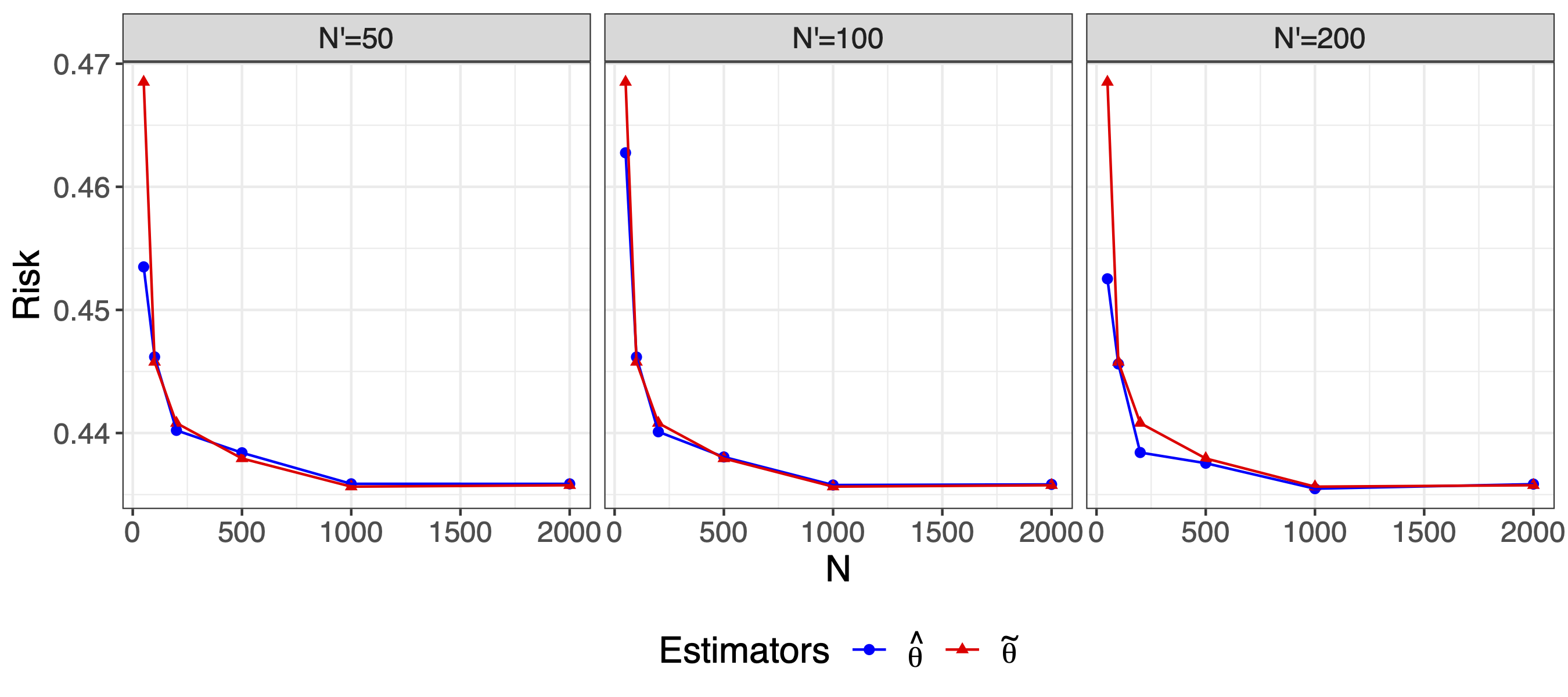}
\caption{Comparison of risk between $\bm{\hat{\theta}}_1$ and $\bm{\tilde{\theta}}_1$ estimators in simulation~\ref{s1}.}
\label{fig:1}
\end{figure}

\begin{figure}[!htb]
\centering
\includegraphics[width=0.9\linewidth]{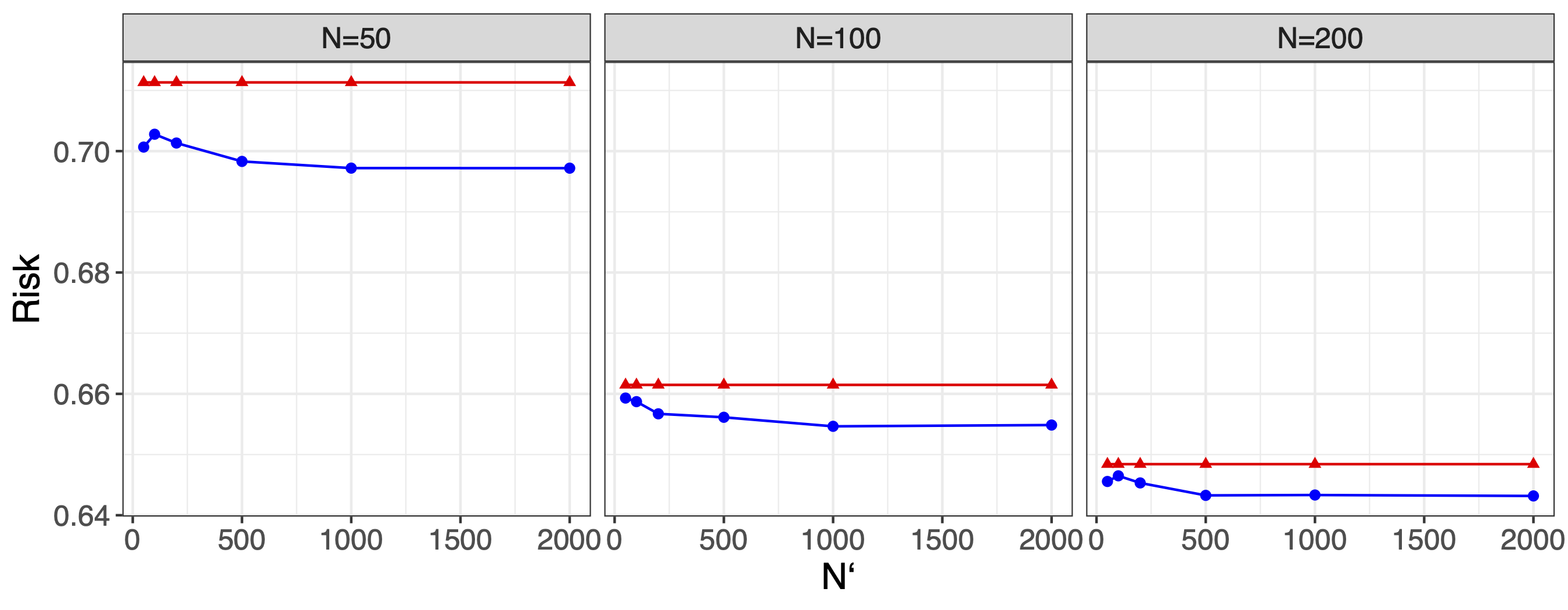}
\includegraphics[width=0.9\linewidth]{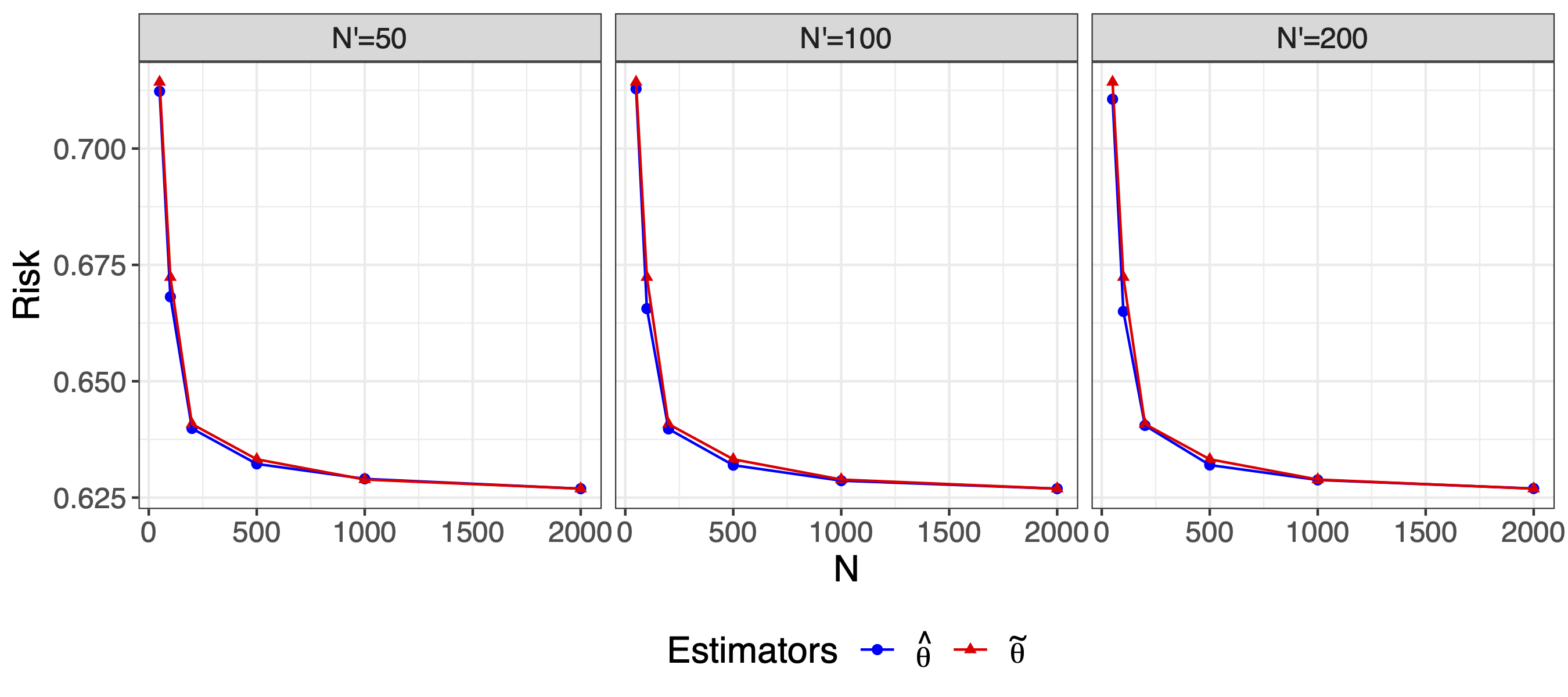}
\caption{Comparison of risk between $\bm{\hat{\theta}}_2$ and $\bm{\tilde{\theta}}_2$ estimators in simulation~\ref{s2}.}
\label{fig:2}
\end{figure}

The results are shown in Figure~\ref{fig:1} and Figure~\ref{fig:2} for disjoint index sets. Overall, as we increase $k$ while fixing other parameters, the risk of both $\hat{\bm \theta}$ and $\Tilde{\bm \theta}$ increases, which is as expected since we aim to capture the variability at a finer level with the same sample size. The first row of the plot for both scenarios represents the case when we fix $N$ and vary $N'$. Note that the sufficient condition for the dominance of $\hat{\bm \theta}$ is not satisfied in~\ref{s1}, and we notice some cases where $\Tilde{\bm \theta}$ has a lower risk when $N'$ is relatively small. In contrast, $\hat{\bm \theta}$ outperforms  $\Tilde{\bm \theta}$ overall in~\ref{s2} even when $N'$ is small. The second row represents the case when we fix $N'$ and vary $N$. The risk decreases drastically as $N$ increases due to the fact that we incorporate more and more observations that are fully classified.

\begin{figure}[!htb]
\centering
\includegraphics[width=0.9\linewidth]{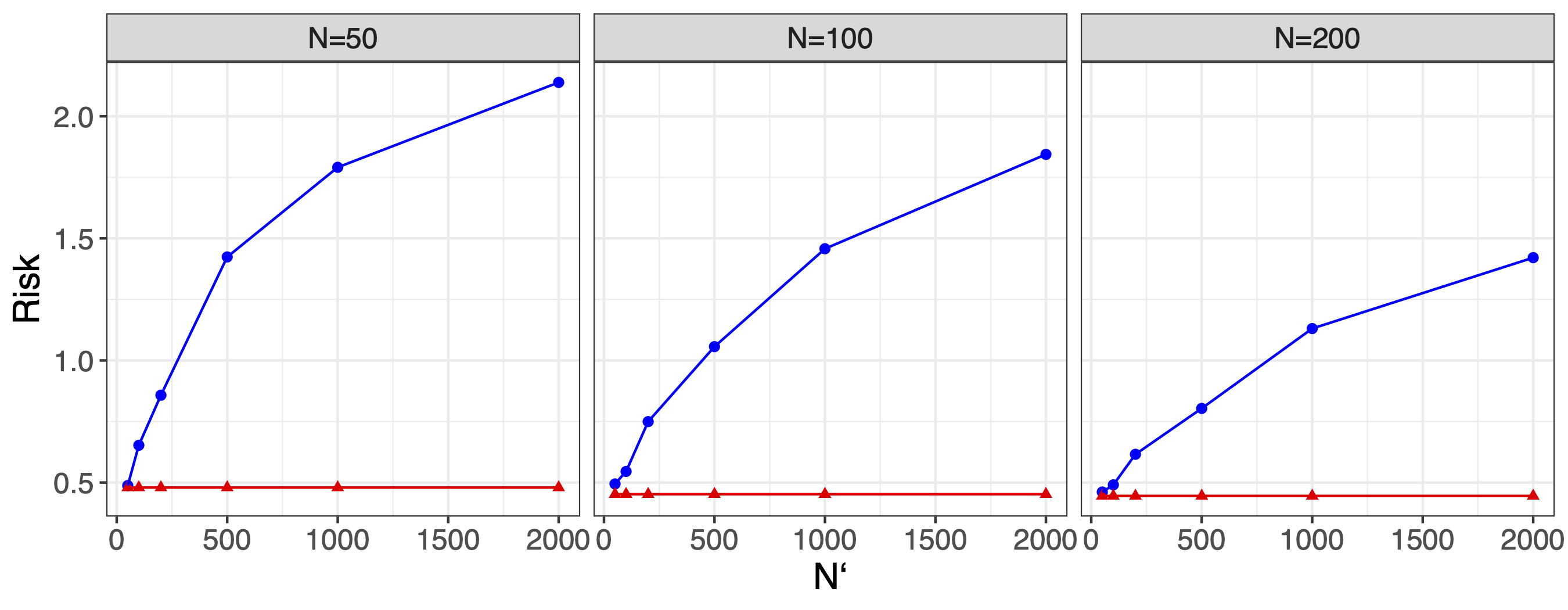}
\includegraphics[width=0.9\linewidth]{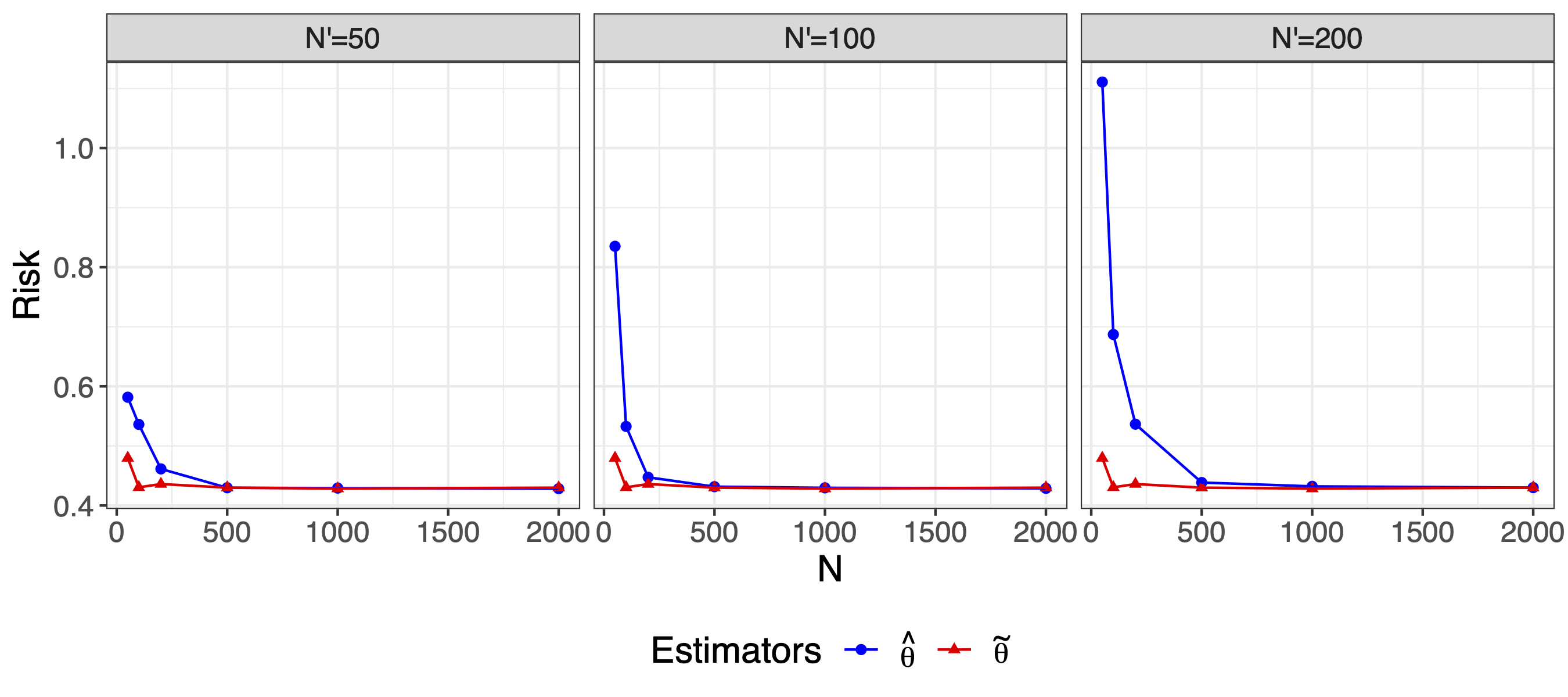}
\caption{Comparison of risk between  $\bm{\hat{\theta}}_1$ and $\bm{\tilde{\theta}}_1$ estimators in simulation~\ref{s3}.}
\label{fig:3}
\end{figure}

\begin{figure}[!htb]
\centering
\includegraphics[width=0.9\linewidth]{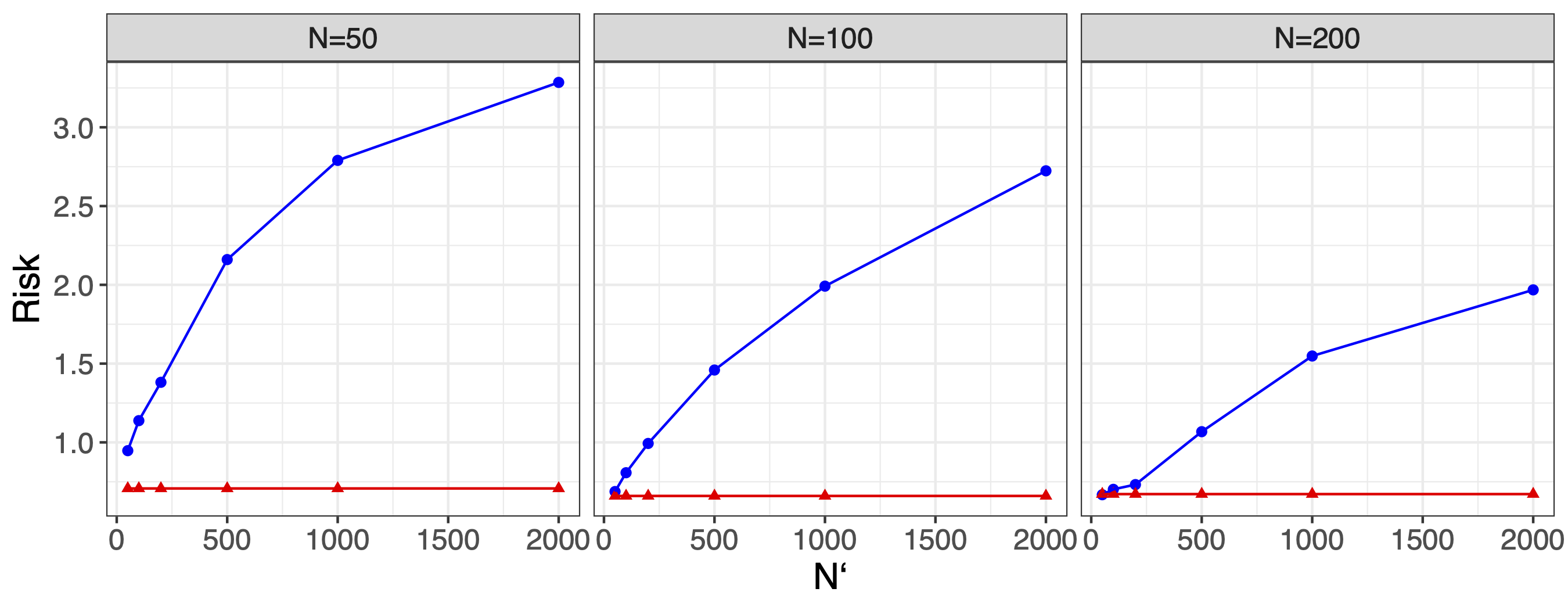}
\includegraphics[width=0.9\linewidth]{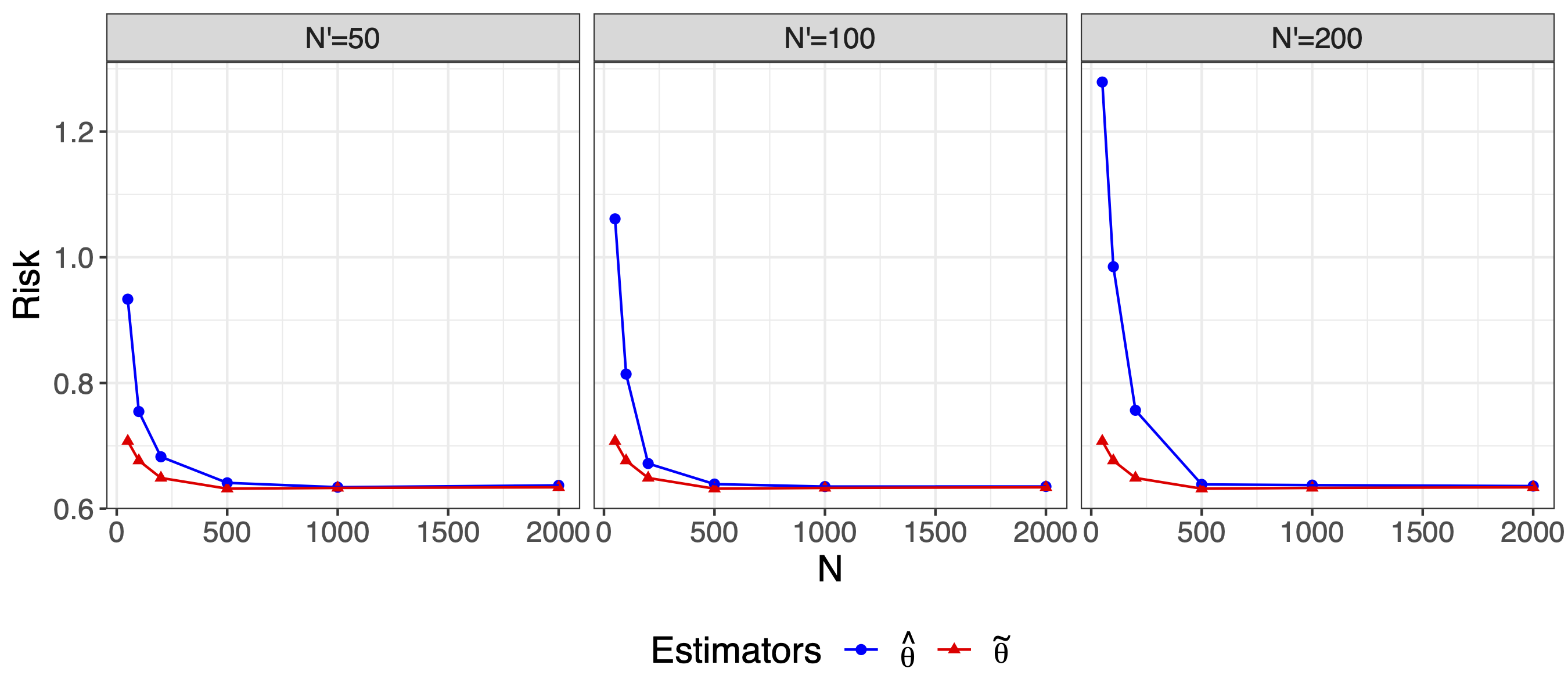}
\caption{Comparison of risk between  $\bm{\hat{\theta}}_2$ and $\bm{\tilde{\theta}}_2$ estimators in simulation~\ref{s4}.}
\label{fig:4}
\end{figure}

Figure~\ref{fig:3} and Figure~\ref{fig:4} display the Bayes risk of $\hat{\bm \theta}_1$ and 
$\hat{\bm \theta}_2$ estimated using Gibbs sampling. Overall, we observe that $\hat{\bm \theta}$ has a higher risk than $\tilde{\bm \theta}$, particularly when $N'$ is large. When $N'$ is fixed, increasing $N$ leads to a decrease in the risk of $\hat{\bm \theta}$, as incorporating more fully classified data reduces the uncertainties introduced by adding $N'$.

We test our data disaggregation approach on two empirical examples, where we utilize data from the Sample Registration System (SRS) and Demographic and Health Surveys (DHS) as the sources of information, respectively. 

\subsection{MCHSS Data of China}

In this section, we provide the first empirical example using MCHSS data~\citep{schumacher2020flexible} obtained through China's sample registration system dedicated to maternal and child health. Over a span of 20 years, from 1996 to 2015, all deaths of children under five years of age residing within the surveillance areas were recorded and grouped into six distinct age categories: 0-6 days, 7-27 days, 1-5 months, 6-11 months, 12-23 months, and 24-59 months, with eight non-overlapping, exhaustive categories of CODs. 

To demonstrate the effectiveness of our disaggregation method under a nested age structure, we create a synthetic dataset based on the MCHSS data collected during the period of $1996$ to $2005$.  In the synthetic dataset, the ages are partially classified into three groups: (i) 0-27 days, (ii) 1-11 months, and (iii) 12-59 months. Our goal is to use the fully classified MCHSS data collected from $2006$ to $2015$ to perform age group disaggregation. 

\begin{table}[!htb]
\parbox[b]{0.48\linewidth}{
\centering
\caption*{True 1996-2005 MCHSS Data}
\vspace{1mm}
\begin{tabular}{l|llllll}
  \hline
 & 1 & 2 & 3 & 4 & 5 & 6 \\ 
  \hline
1 & 3550 & 638 & 138 & 9 & 2 & 0 \\ 
  2 & 4398 & 183 & 24 & 0 & 0 & 0 \\ 
  3 & 1783 & 568 & 732 & 373 & 264 & 270 \\ 
  4 & 95 & 78 & 459 & 224 & 268 & 473 \\ 
  5 & 732 & 231 & 463 & 176 & 584 & 1333 \\ 
  6 & 39 & 92 & 501 & 361 & 270 & 196 \\ 
  7 & 1227 & 991 & 1671 & 570 & 428 & 328 \\ 
  8 & 1051 & 722 & 491 & 269 & 230 & 261 \\ 
   \hline
\end{tabular}
}
\parbox[b]{0.48\linewidth}{
\centering
\caption*{Predicted 1996-2005 MCHSS Data}
\vspace{1mm}
\begin{tabular}{l|llllll}
  \hline
 & 1 & 2 & 3 & 4 & 5 & 6 \\ 
  \hline
1 & 3474 & 714 & 145 & 2 & 1 & 1 \\ 
  2 & 4258 & 323 & 23 & 1 & 0 & 0 \\ 
  3 & 1659 & 692 & 740 & 365 & 291 & 243 \\ 
  4 & 101 & 72 & 454 & 229 & 274 & 467 \\ 
  5 & 572 & 391 & 471 & 168 & 667 & 1250 \\ 
  6 & 26 & 105 & 536 & 326 & 321 & 145 \\ 
  7 & 1094 & 1124 & 1635 & 606 & 435 & 321 \\ 
  8 & 1059 & 714 & 515 & 245 & 243 & 248 \\ 
   \hline
\end{tabular}
}
\vspace{4mm}
\caption{Comparison of true and disaggregated MCHSS data using Bayesian age reconciliation for the years 1996-2005.}
\label{tab:mchss-pred}
\end{table}

We evaluate the effectiveness of our proposed Bayesian age reconciliation method by comparing its predicted results to the actual MCHSS data from 1996 to 2005, as shown in Table~\ref{tab:mchss-pred}. We also compare the accuracy of our method to the data integration by random assignment approach.
Our findings demonstrate that our method performs well, achieving a prediction accuracy of $93.0\%$ while only $52.6\%$ of the observations are classified correctly using the random assignment. These results highlights our method's ability to preserve the joint COD-age distribution information with high accuracy. 

\begin{figure}[!htb]
\centering
\vspace{1mm}
\caption*{True 1996-2005 MCHSS Data}

\includegraphics[width = 0.7\linewidth]{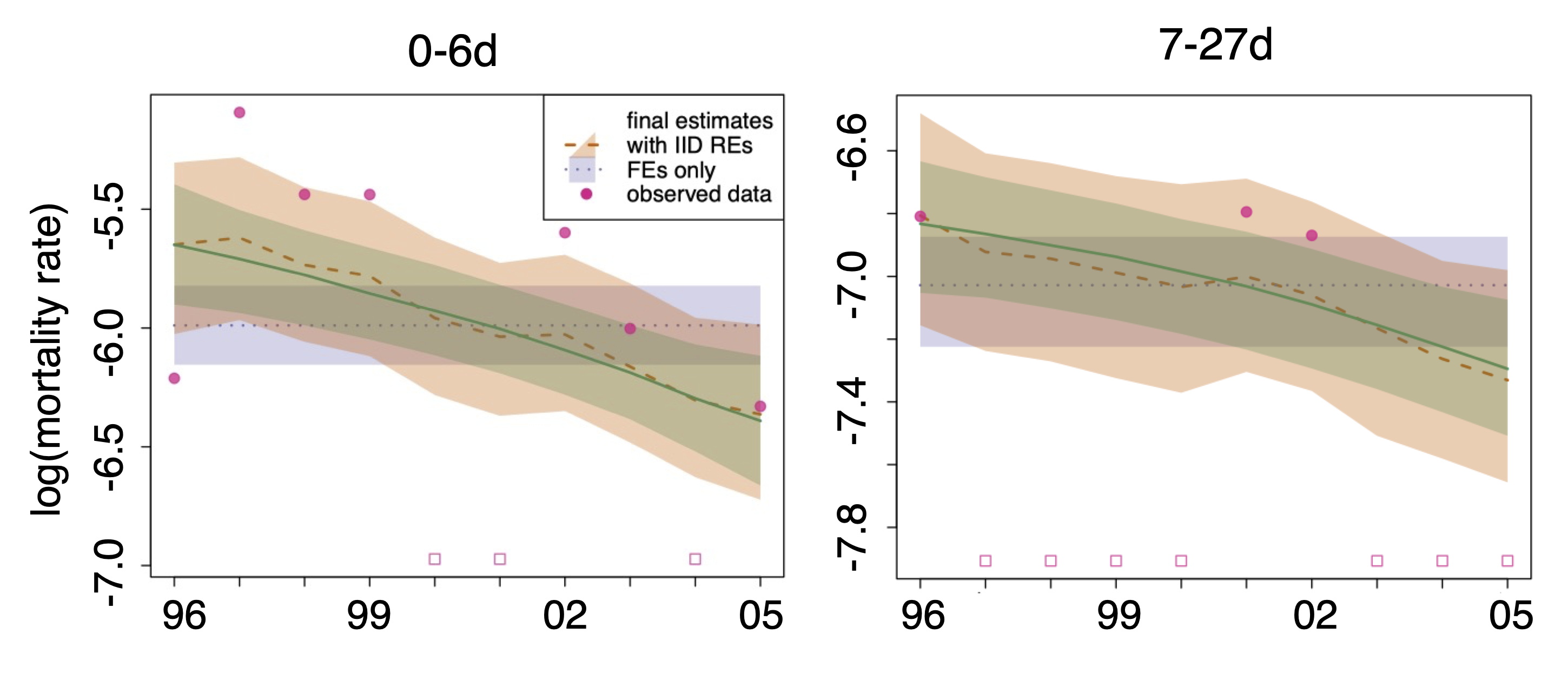}
\vspace{1mm}
\caption*{Predicted 1996-2005 MCHSS 
Data}

\includegraphics[width = 0.7\linewidth]{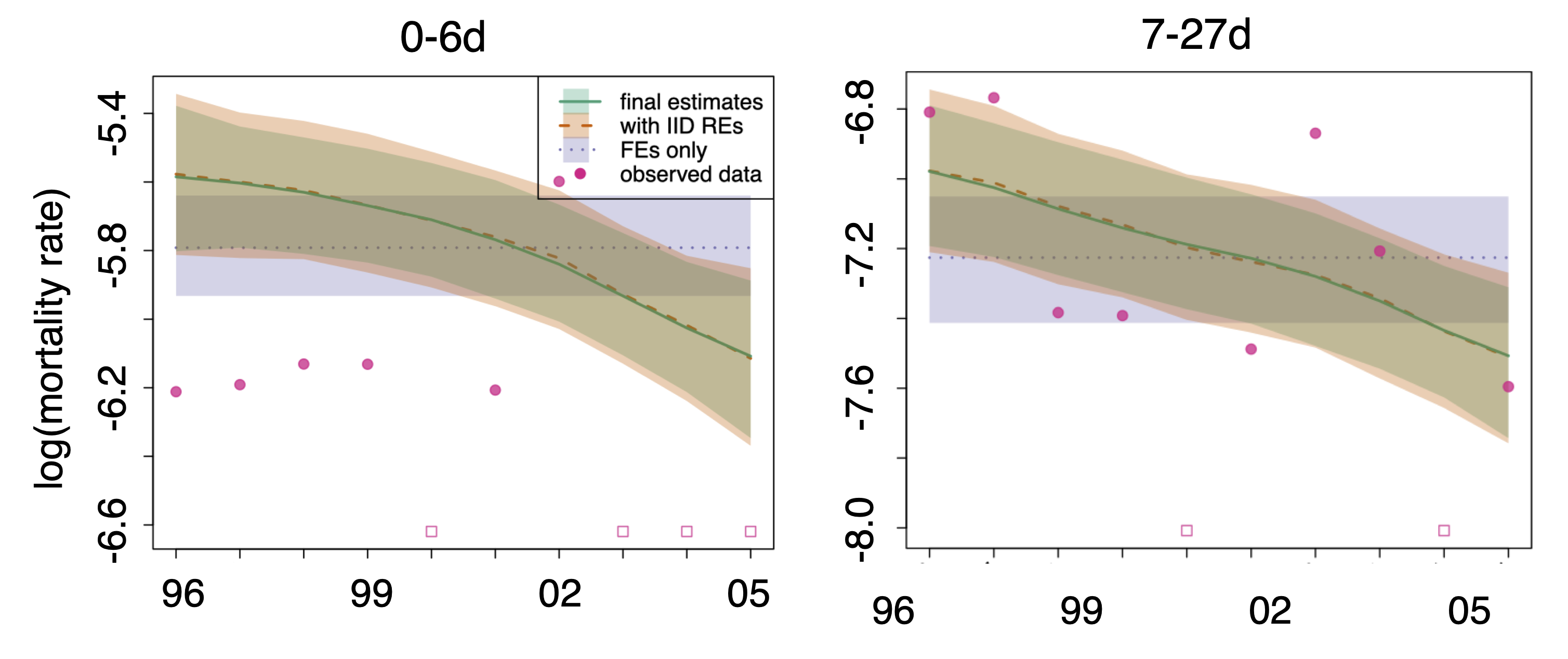}
\vspace{1mm}
\caption{Estimation of log mortality rates for non-communicable diseases in the east urban region using MCHSS data. The plot shows empirical data, estimated posterior medians, and posterior 80\% intervals. Combinations with zero deaths are indicated by an open square. }
\label{fig:logmx-compare-east-urban}
\end{figure}

To examine the impact of utilizing the disaggregated data on the estimation of age- and cause-specific child mortality, we fit separate Bayesian models to the true and predicted 1996-2005 MCHSS data, as proposed by~\cite{schumacher2020flexible}. Figures~\ref{fig:logmx-compare-east-urban} present the posterior medians and $80\%$ intervals for the estimated log mortality rates in each age group over the period of 1996-2005 based on the true and predicted data in selected causes and ages. Additionally, we provide posterior median and $80\%$ intervals for the estimated log mortality rates in models with fixed effects only and models with added random effect error terms, as previously discussed in~\cite{schumacher2020flexible}.

\begin{figure}[!htb]
\centering
\caption*{True 1996-2005 MCHSS Data }

\includegraphics[width=0.8\linewidth]{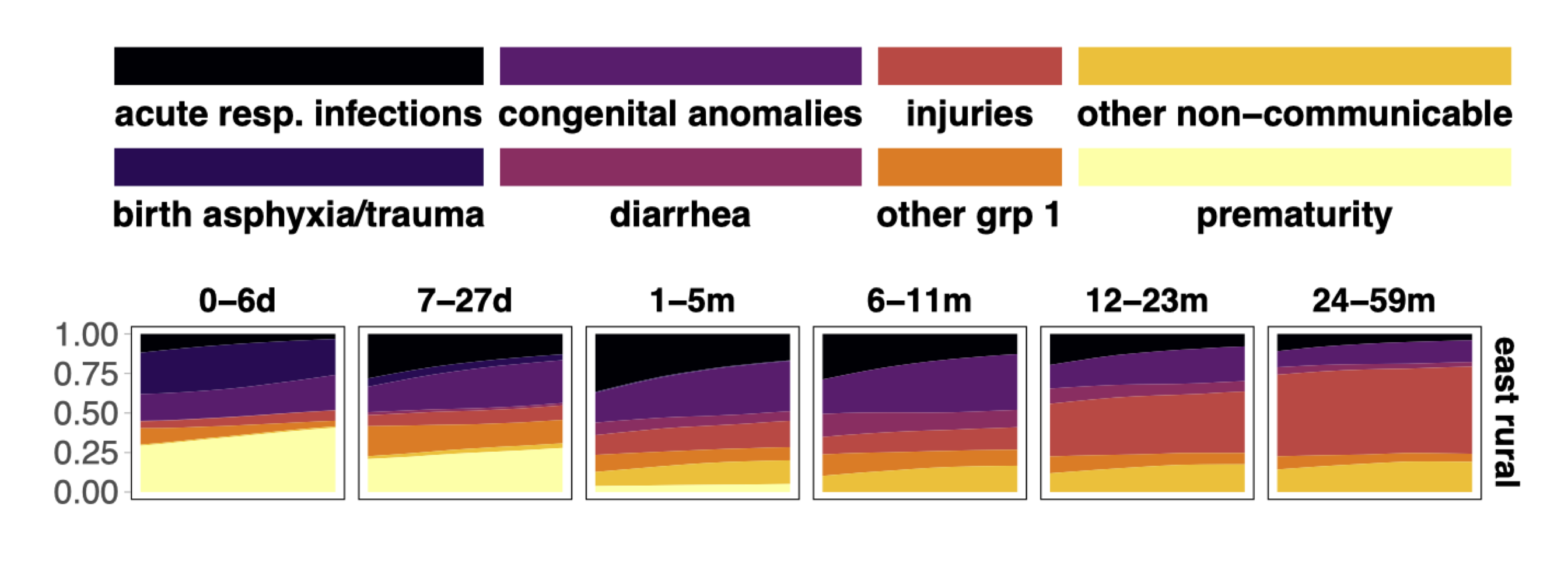}

\caption*{Predicted 1996-2005 MCHSS Data}

\includegraphics[width=0.8\linewidth]{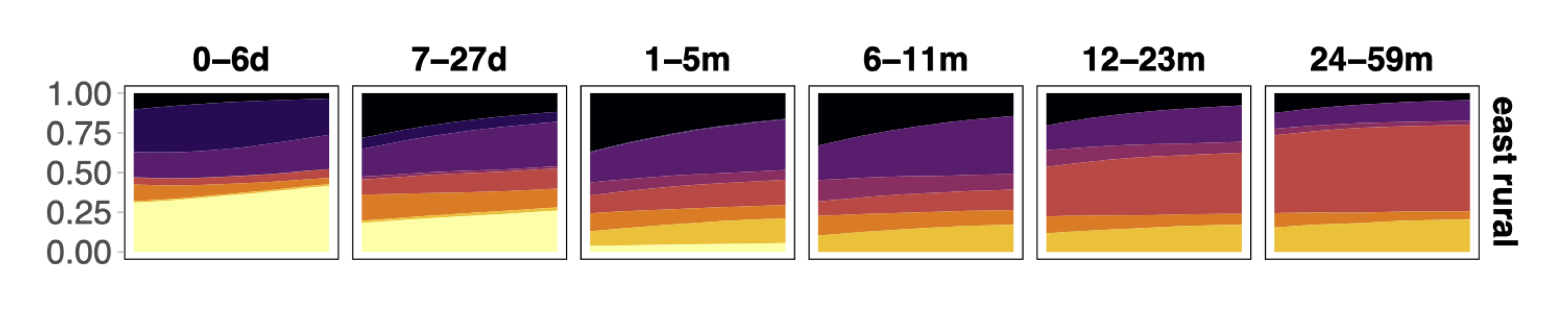}
\vspace{4mm}
\caption{Comparisons of estimated CSMFs between models based the true MCHSS data and the estimated MCHSS data for selected regions, showing agreement in temporal trends and estimated CSMFs.}
\label{fig:csmf-compare}
\end{figure}

Figure~\ref{fig:logmx-compare-east-urban} shows the results of the analysis of non-communicable diseases in the 0-6 days and 7-27 days age groups. Both the true and predicted data fit the models well, but some discrepancies are observed. Specifically, the estimated log mortality rates in the 0-6 days age group are consistently higher when using the predicted data compared to the true data. This discrepancy may be due to some deaths being incorrectly classified in the 7-27 days age group during the disaggregation process. However, the model fitted to the predicted data effectively captures the overall age-specific time trend and benefits from borrowing strength from other age strata.

Furthermore, in Figure~\ref{fig:csmf-compare}, the estimated cause-specific mortality fractions (CSMFs) in the selected region are compared between the model fit of the real MCHSS data and the predicted data. This comparison demonstrates that the temporal trends and the estimated CSMFs from the model fit of the predicted data are in agreement with those of the true data.

\subsection{BDHS Data}

The BDHS data was collected through VA questionnaires in 2011, as well as in 2017-2018. Physical reviews were performed to determine the COD for each individual, with 11 common CODs being recorded. The ages at death for each individual were also documented. To demonstrate  the effectiveness of our proposed method under a non-nested age structure, we aggregated the 2011 data into non-standard age categories: (i) 0-3 months, (ii) 4-11 months, and (iii) 12-59 months. For the 2017-2018 data, we created standardized age-disaggregated data with six age groups, as listed in Table~\ref{tab:age}.

Table~\ref{tab:bdhs-pred} presents the comparison between the actual and predicted 2011 BDHS data, which shows overall satisfactory results, albeit with a few instances of misclassification in the non-nested age categories. Our proposed method achieved a slightly improved prediction accuracy of $70.4\%$, compared to $54.9\%$ with random assignment.

\begin{table}[!htb]
\parbox[b]{0.48\linewidth}{
\centering
\caption*{True 2011 BDHS data}
\vspace{1mm}
\begin{tabular}{l|llllll}
  \hline
 & 1 & 2 & 3 & 4 & 5 & 6 \\ 
  \hline
  1 &   4 &   4 &   0 &   0 &   0 &   0 \\ 
  2 &   1 &   3 &   0 &   0 &   0 &   0 \\ 
  3 &   0 &   0 &   0 &   1 &  13 &  12 \\ 
  4 &  55 &   2 &   0 &   0 &   0 &   0 \\ 
  5 &   9 &   0 &   0 &   0 &   0 &   0 \\ 
  6 &   0 &   0 &   4 &   2 &   4 &   0 \\ 
  7 &  22 &  21 &  39 &  14 &   6 &   7 \\ 
  8 &   5 &   5 &   0 &   0 &   0 &   0 \\ 
  9 &  26 &   5 &   0 &   0 &   0 &   0 \\ 
  10 &  44 &  14 &   0 &   0 &   0 &   0 \\ 
  11 &   0 &   0 &   1 &   1 &   0 &   0 \\ 
   \hline
\end{tabular}
}
\parbox[b]{0.48\linewidth}{
\centering

\caption*{Predicted 2011 BDHS data}
\vspace{1mm}
\begin{tabular}{l|llllll}
  \hline
 & 1 & 2 & 3 & 4 & 5 & 6  \\ 
  \hline
1 & 4 & 3 & 1 & 0 & 0 & 0 \\ 
  2 & 2 & 1 & 1 & 0 & 0 & 0 \\ 
  3 & 0 & 0 & 0 & 1 & 13 & 12 \\ 
  4 & 55 & 2 & 0 & 0 & 0 & 0 \\ 
  5 & 8 & 1 & 0 & 0 & 0 & 0 \\ 
  6 & 0 & 0 & 4 & 2 & 2 & 2 \\ 
  7 & 23 & 13 & 57 & 3 & 2 & 11 \\ 
  8 & 7 & 3 & 0 & 0 & 0 & 0 \\ 
  9 & 26 & 4 & 1 & 0 & 0 & 0 \\ 
  10 & 31 & 23 & 4 & 0 & 0 & 0 \\ 
  11 & 0 & 1 & 1 & 0 & 0 & 0 \\ 
   \hline
\end{tabular}
}
\vspace{4mm}
\caption{Evaluating proposed method for non-standard age categories using 2011 BDHS data: comparison of actual and predicted results demonstrates promising performance with some misclassification observed.}
\label{tab:bdhs-pred}
\end{table}

\section{Discussion}

In this work, we extend the existing literature on simultaneous estimation of multinomial probabilities to a broader and more versatile setting. We provide theoretical support and numerical studies, demonstrating how the integration of partially classified data can lead to improved Bayesian estimators of multinomial probabilities. Our proposed age reconciliation method is based on this approach and has been tested through simulation studies using observed, disaggregated data. The results show that our method is promising and offers a novel perspective on mitigating age inconsistencies in ACSU5M studies.

Our proposed age reconciliation method is a promising first step in tackling the problem of age inconsistencies in ACSU5M studies. However, there are several future directions for research that could further address this issue. First, the CODs are commonly assigned and analyzed by statistical algorithms. For example, several Bayesian methods have been developed to infer CODs based on verbal autopsies~\citep[e.g,][]{mccormick2016probabilistic,kunihama2020bayesian,li2021bayesian,wu2021tree} and estimate the population-level cause specific mortality fractions~\citep[e.g,][]{serina2015improving,byass2019integrated,moran2021}. However, it has been shown that considerable uncertainties exist in the classification in these models. One possible extension of our proposed method is to account for misclassifications in both CODs and age groups through the use of joint misclassification matrices in the model parameter estimations. This can be done by extending Bayesian hierarchical models such as the one proposed by~\cite{mulick2021bayesian}. Our proposed method can extend the approach by accounting for misclassifcations from both CODs and age groups through of use of joint misclassification matrices in the model parameter estimations. Second, although our proposed method offers improved estimators under certain conditions, it fails to provide information theoretic metrics that quantify the effects of data disaggregation. One possible extension of the current approach is to use Dampster-Shafer inference~\citep{dempster1976mathematical} procedures as an alternative approach for multinomial inference~\citep{lawrence}. This approach can incorporate epistemic uncertainty and have the potential to quantify the effects of adding partial-classified multinomial data as adversarial attacks. Overall, we believe that these future research directions can further improve the accuracy and reliability of age reconciliations in ACSU5M studies.

\bibliographystyle{unsrt}
% \bibliography{references}  %%% Uncomment this line and comment out the ``thebibliography'' section below to use the external .bib file (using bibtex) .


\pagebreak

\appendix
\renewcommand{\thesection}{\Alph{section}}
\renewcommand{\thesubsection}{S.\arabic{subsection}}

\section*{\myformat{Supplementary Material to ``Bayesian Age Category Reconciliation for Age- and Cause-specific Under-five Mortality Estimates''} }
\beginsupplement

\subsection{Proof of Lemma 1}

Let $\bm\rho_{A_j} = (\frac{\theta_i}{\tau_j},i\in A_j)^T$, $j = 0, \dots, m$ be the conditional probabilities of each individual cell within each group $A_j$ and $\bm \eta = (\bm \tau, \bm \rho_{A_j}, j = 0,\dots, m)$
 The probability mass function of $(\bm X, \bm Y) = (X_1,X_2,\dots, X_k, Y_{A_1}, \dots, Y_{Am})$ is
\begin{align*}
    f(\bm X, \bm Y|\bm\eta) &= \left(\frac{N!}{\prod_{i=1}^k X_i!} \prod_{i=1}^k \theta_i^{X_i}\right) \frac{N'!}{\prod_{j=0}^m Y_{A_j}!}\prod_{j=0}^m \tau_j^{Y_{A_j}}\\
    &=  \frac{N!}{\prod_{i=1}^k X_i!}\frac{N'!}{\prod_{j=0}^m Y_{A_j}!} \left\{\prod_{r\in A_0}(\tau_0\rho_{A_0}^{(r)})^{X_r}\times \prod_{r\in A_1}(\tau_1\rho_{A_1}^{(r)})^{X_r}\times \dots \times \prod_{r\in A_m}(\tau_m\rho_{A_m}^{(r)})^{X_r}\right\}\prod_{j=0}^m \tau_j^{Y_{A_j}} \\
    &= \frac{N!}{\prod_{i=1}^k X_i!}\frac{N'!}{\prod_{j=0}^m Y_{A_j}!} \left(\prod_{j=0}^m \prod_{r\in A_j}(\tau_j \rho_{A_j}^{(r)})^{X_r}\right)\prod_{j=0}^m \tau_j^{Y_{A_j}}\\
    &= \frac{N!}{\prod_{i=1}^k X_i!}\frac{N'!}{\prod_{j=0}^m Y_{A_j}!}\left(\prod_{j=0}^m \tau_j^{\sum_{r\in A_j}X_r + Y_{A_j}}\right)\prod_{j = 0}^m\prod_{r\in A_j}(\rho_{A_j}^{(r)})^{X_r}\\
    &= \frac{N!}{\prod_{i=1}^k X_i!}\frac{N'}{\prod_{j=0}^m Y_{A_j}!}\left(\prod_{j=0}^m \tau_j^{X_{A_j} + Y_{A_j}}\right)\prod_{j = 0}^m\prod_{r\in A_j}(\rho_{A_j}^{(r)})^{X_r}
\end{align*}
where $X_{A_j} = \sum_{r\in A_j} X_r$.

\begin{align*}
   \log \{\prod_{j = 0}^m\prod_{r\in A_j}(\rho_{A_j}^{(r)})^{X_r}\} &= \log \{ \prod_{r\in A_0}(\rho_{A_0}^{(r)})^{X_r}\times \prod_{r\in A_1}(\rho_{A_1}^{(r)})^{X_r} \times \dots \times \prod_{r\in A_m}(\rho_{A_m}^{(r)})^{X_r}\}\\
   &= \sum_{j=0}^m  \log \prod_{r\in A_j}(\rho_{A_j}^{(r)})^{X_r}\\
   &= \sum_{j=0}^m \sum_{r\in A_j} X_r\log(\rho_{A_j}^{(r)})\\
   &= \sum_{j=0}^m \sum_{r\in A_j\backslash j^{0}} X_r\log(\rho_{A_j}^{(r)}) + X_{j^{0}}\log_{\rho_{A_j}^{j^{0}}}\\
   % &= \sum_{j=0}^m \sum_{r\in A_j\backslash j^{0}} X_r\log(\rho_{A_j}^{(r)}) + (1 - \sum_{r\in A_j\backslash j^{0}} X_r\log(\rho_{A_j}^{(r)}))
\end{align*}

where $j^{0} = A_j^{(0)}$ is the first item of $A_j$. 

For $r \in A_j\backslash j_0$, $j = 0,\dots, m$:

\[\frac{\partial}{\rho_{A_j}^{(r)}} \log f(\bm X, \bm Y|\bm \eta) = \frac{X_{r}}{\rho_{A_j}^{(r)}} - \frac{X_{j^0}}{\rho_{A_j}^{(j^0)}}\]

For $r, r' \in  A_j\backslash j_0 $ and $j = 0, \dots, m$: 

\[\frac{\partial^2}{(\partial\rho_{A_j}^{(r)})(\partial\rho_{A_j}^{(r')})} \log f(\bm X, \bm Y|\bm \eta) =  
\begin{cases}
-\frac{X_r}{(\rho_{A_j}^{(r)})^2} - \frac{X_{j^0}}{(\rho_{A_j}^{j^0})^2} & \mbox{ if } r = r' \\
-\frac{X_{r}}{(\rho_{A_j}^{(r)})^2} & \mbox{ if } r\neq r'
\end{cases}\]

Let $A_p = \{p_1, \dots, p_{n_p}\}$, $p = 0, \dots, m$. For the $p$-th group, the conditional vector of the cells within the group is $\rho_{A_p} = (\frac{\theta_{p_1}}{\tau_p}, \dots, \frac{\theta_{p_{n_p}}}{\tau_p})$. Consider the Jeffrey's prior, we have that 

The Jefferys prior is 

\begin{align*}
\pi(\bm\eta)d\bm\eta &\propto \sqrt{\left| \mathbb{E}_{\bm\theta}\left[ \mbox{diag}\left(\frac{X_{A_1} + Y_{A_1}}{\tau_1^2},\dots, \frac{X_{A_m} + Y_{A_m}}{\tau_m^2}\right) + \bm i^{(m)}(\bm i^{(m)})^T\frac{X_{A_0}+Y_{A_0}}{\tau_0^2}\right] \right|}\\
& \times \prod_{j = 0}^m \sqrt{\left|\mathbb{E}_{\bm \theta}\left[\mbox{diag}\left(\frac{X_{r}}{(\rho_{A_j}^{(r)})^2}\right)_{r\in A_j\backslash j_0}+ \bm i^{(n_j-1)}(\bm i^{(n_j-1)})^T\frac{X_{j^0}}{(\rho_{A_j}^{(j^0)})^2}\right] \right|}\\
& \propto \sqrt{\left|\mbox{diag}\left(\frac{1}{\tau_1}, \dots, \frac{1}{\tau_m}\right)  +  \bm i^{(m)}(\bm i^{(m)})^T \frac{1}{\tau_0} \right|} \times \left(\prod_{j=1}^m(\sqrt{\tau_j})^{n_j}\right)\\
&\times \prod_{j=0}^m \sqrt{\left| \mbox{diag} \left(\frac{1}{\rho_{A_j}^{(r)}}\right)_{r\in A_j\backslash j^0} + \bm i^{(n_j-1)}(\bm i^{(n_j-1)})^T \frac{1}{\rho_{A_j}^{(j^0)}} \right|}\\
&= \sqrt{\left| \mbox{diag}\left(\frac{1}{\tau_1}, \dots, \frac{1}{\tau_m}\right) \right| \left[1 + \frac{ (\bm i^{(m)})^{T}}{\sqrt{\tau_0}}\left\{\mbox{diag} \left(\frac{1}{\tau_1}, \dots, \frac{1}{\tau_m}\right)\right\}^{-1}\frac{\bm i^{(m)}}{\sqrt{\tau_0}}\right]}\\
&\times \left(\prod_{j=1}^m \tau_j^{n_j/2}\right) \prod_{j=0}^m \sqrt{\left| \mbox{diag}\left(\frac{1}{\rho_{A_j}^{(r)}}\right)_{r\in A_j\backslash j^0}  \right|\left[1 + \frac{(\bm i^{(n_j-1)})^T}{\sqrt{\rho_{A_j}^{(j^0)}}}\left\{\mbox{diag}\left(\frac{1}{\rho_{A_j}^{(r)}}\right)_{r\in A_j\backslash j^0}\right\}^{-1}\frac{\bm i^{(n_j-1)}}{\sqrt{\rho_{A_j}^{(j^0)}}}\right] }\\
&= \left\{\prod_{j=0}^m \tau_i^{n_j/2 - 1}\right\}\prod_{j=0}^m\prod_{r\in A_j}(\rho_{A_j}^{(r)})^{1/2 - 1}
\end{align*}

The Bayes estimators under the KL loss are the posterior means of $\bm\theta = (\bm \tau, \bm \rho_{A_j}, j = 0, \dots, m)$. Therefore, with respect to the Jeffreys prior and the direct observations $(\bm X, \bm Y)$, the Bayes estimator is given as 
\begin{align*}
\hat{\bm \theta} &= (\theta_i)_{i=1}^{k} \mbox{, where for }i\in A_j,\\\hat{\theta_i} &= \mathbb{E}(\theta_i|\bm X, \bm Y) = \mathbb{E}(\frac{\theta_i}{\sum_{r\in A_j}\theta_r}|\bm X, \bm Y)\mathbb{E}(\sum_{r\in A_j}\theta_r|\bm X, \bm Y)\\
&=  \mathbb{E}(\rho_{A_j}^{(i)}|\bm X, \bm Y)\mathbb{E}(\tau_j|\bm X, \bm Y)
\end{align*}

The closed form posterior mean for $\tau_j$ and $\rho_{A_j}^{(i)}$ can be obtained as 
\[\mathbb{E}\left(\rho_{A_j}^{(i)}\bigg|\bm X, \bm Y\right) = \mathbb{E}\left(\frac{\theta_i}{\sum_{r\in A_j}\theta_r}\bigg|\bm X, \bm Y\right) = \frac{1/2 + x_i}{\sum_{r\in A_j}x_r + n_j/2}\]
\begin{align*}
\mathbb{E}\left(\tau_j|\bm X, \bm Y\right) = \mathbb{E}\left (\sum_{r\in A_j}\theta_r\bigg|\bm X, \bm Y \right) &= \frac{y_{A_j} + \sum_{r\in A_j}x_r + n_j/2}{\sum_{l=0}^m \{y_{A_l} + \sum_{r\in A_l}(1/2 + x_{r})\}} \\
&= \frac{y_{A_j} + \sum_{r\in A_j}x_r + n_j/2}{N + N' + n_j(m+1)/2}
\end{align*}

Therefore,
\[\hat{\theta_i} = \frac{1/2 + x_i}{\sum_{r\in A_j}x_r + n_j/2}\frac{y_{A_j} + \sum_{r\in A_j}x_r + n_j/2}{N + N' + n_j(m+1)/2}\]

We can further show that the posterior of $(\bm \eta)$ with respect to a prior distribution for $\bm \theta$ being any Dirichlet distribution with parameters $\bm \alpha = (\alpha_1, \alpha_2, \dots, \alpha_k)$ admits a tractable representation. 

\begin{align*}
\hat{\theta}_i = \frac{\alpha_i + x_i}{\sum_{r\in A_j}(x_r + \alpha_r)}  \frac{y_{A_j} + \sum_{r\in A_j}(\alpha_r + x_r)}{\sum_{l=0}^m\{y_{A_l} + \sum_{r\in A_l}(\alpha_r + x_r)\}}
\end{align*}

\subsection{Proof of Lemma 2}

We would like to compare the risk functions of $\tilde{\bm \theta}$ and $\hat{\bm \theta}$. 
\[\Delta_{\bm\theta}(N, N') = \mathbb{E}_{\bm\theta}[L(\hat{\bm \theta}, \bm\theta)] - \mathbb{E}_{\bm\theta}[(L(\tilde{\bm \theta},\bm \theta)]\]

For notation simplicity, let $x_{A_j} = \sum_{r\in A_j}(x_r)$, $\theta_{A_j} = \sum_{i\in A_j} \theta_i$, $\alpha_{A_j} = \sum_{r\in A_j}\alpha_r$, and $\alpha_S = \sum_{c=1}^k \alpha_c$.

\begin{align*}
    \Delta_{\bm\theta}(N, N') &=  \mathbb{E}_{\bm\theta}\left[\sum_{j=0}^m\sum_{i\in A_j} \theta_i \log \frac{\tilde{\theta}_i}{\hat{\theta}_i}\right]\\
    &= \mathbb{E}_{\bm \theta}\left[\sum_{j=0}^m\sum_{i\in A_j}\theta_i \log\left\{\frac{x_i + \alpha_i}{\sum_{c=1}^k \alpha_c + x_c}\times \frac{\sum_{r\in A_j}(x_r + \alpha_r)} {\alpha_i + x_i}\frac{\sum_{l=0}^m\{y_{A_l} + \sum_{r\in A_l}(\alpha_r + x_r)\}}{y_{A_j} + \sum_{r\in A_j}(\alpha_r + x_r)} \right\} \right]\\
    &= \mathbb{E}_{\bm \theta}\left[\sum_{j=0}^m\sum_{i\in A_j}\theta_i \log\left\{\frac{N' + N + \sum_{l=0}^m\sum_{r\in A_l}(\alpha_r)}{N +\sum_{c=1}^k \alpha_c }\times \frac{\sum_{r\in A_j}(x_r + \alpha_r)}{y_{A_j} + \sum_{r\in A_j}(\alpha_r + x_r)} \right\} \right]\\
    &= \mathbb{E}_{\bm \theta}\left[\sum_{j=0}^m\sum_{i\in A_j}\theta_i \log\left\{\frac{N' + N + \sum_{l=0}^m\alpha_{A_l}}{N +\alpha_S }\times \frac{x_{A_j} + \alpha_{A_j}}{y_{A_j} +x_{A_j} + \alpha_{A_j} } \right\} \right]\\
    &= \mathbb{E}_{\bm \theta}\left[\log \frac{N' + N + \alpha_S}{N +\alpha_S }   +\sum_{j=0}^m \theta_{A_j}\log\left\{\frac{x_{A_j} + \alpha_{A_j}}{y_{A_j} +x_{A_j} + \alpha_{A_j} }\right\}\right]\\
    &= \sum_{u=1}^{N'}\Delta_{\bm \theta}(N + u - 1, 1)
\end{align*}

\subsection{Proof of Lemma 3}

Without loss of generality, we consider $\Delta_{\bm \theta}(N, 1)$. Let $(Z_{A_j})_{j=0}^m \sim \mbox{Multi}(1, (\theta_{A_j})_{j=0}^m)$ be an independent set of multinomial variables. 
\begin{align*}
    \Delta_{\bm \theta}(N, 1) &=  \log \frac{1 + N + \alpha_S}{N +\alpha_S } + \mathbb{E}_{\bm \theta}\left[\sum_{j=0}^m \theta_{A_j}\log\left\{\frac{x_{A_j} + \alpha_{A_j}}{z_{Aj} +x_{A_j} + \alpha_{A_j} }\right\}\right]\\
    &= \log \frac{1 + N + \alpha_S}{N +\alpha_S } + \mathbb{E}_{\bm \theta}\left[\sum_{j=0}^m \theta_{A_j}\log\left\{\frac{x_{A_j} + \alpha_{A_j}}{z_{A_j}+x_{A_j} + \alpha_{A_j} }\right\}\bigg| Z_{A_j} = 1\right]\mathbb{P}(Z_{A_j}=1) \\& +\mathbb{E}_{\bm \theta}\left[\sum_{j=0}^m \theta_{A_j}\log\left\{\frac{x_{A_j} + \alpha_{A_j}}{z_{A_j}+x_{A_j} + \alpha_{A_j} }\right\}\bigg| Z_{A_j} = 0\right]\mathbb{P}(Z_{A_j}=0)\\
    &=  \log \frac{1 + N + \alpha_S}{N +\alpha_S } + \mathbb{E}_{\bm \theta}\left[\sum_{j=0}^m \theta_{A_j}^2\log\left\{\frac{x_{A_j} + \alpha_{A_j}}{1 +x_{A_j} + \alpha_{A_j} }\right\}\right]\\
    & = \log\left\{ 1 + \frac{1}{N +\alpha_S }\right\} + \mathbb{E}_{\bm \theta}\left[\sum_{j=0}^m \theta_{A_j}^2\log\left\{1 - \frac{1}{1 +x_{A_j} + \alpha_{A_j} }\right\}\right]
\end{align*}
where the second equality follows because $Z_{A_j}\sim \mbox{Bern}(\theta_{A_j})$, $\mathbb{P}(Z_{A_j} = 1) = \theta_{A_j}$

Define the function $G: (0,1) \to (0, \infty)$ by:
\[G(\theta) = \theta^2 \mathbb{E}\left[-\log \left\{  1- \frac{1}{1 + X(\theta) + \alpha_{A_j}}  \right\} \right],\mbox{  }\theta\in (0,1)\]
where $X(\theta) \sim \mbox{Bin}(N, \theta)$ for $\theta\in (0, 1)$.

Then we have 
\[\Delta_{\bm \theta}(N, 1)  =  \log\left\{ 1 + \frac{1}{N +\alpha_S }\right\} - \sum_{j=0}^m G(\theta_{A_j}) \]

If we can prove the condition required for $G$ to be convex, then the maximum is attained. 

We first prove the following results:
\begin{enumerate}
    \item Suppose $X\sim \mbox{Binom}(n,\theta)$, and let $h(X)$ be a function of $X$. Then we can show that 
\[\frac{\partial}{\partial \theta }\mathbb{E}[h(X)] = \frac{1}{\theta}\mathbb{E}[X\left\{h(X) - h(X-1)\right\}]\]

\textit{Proof}:
Using the definition of the expectation of binomial random variable, and lemma of Johnson(1987), we have that
\begin{align*}
    \frac{\partial}{\partial \theta }\mathbb{E}[h(X)] &= \mathbb{E}\left\{\left(\frac{X}{\theta} - \frac{n-X}{1-\theta}\right)h(X)\right\}\\
    &=\frac{1}{\theta(1-\theta)}\mathbb{E}\left\{\left(X-n\theta\right)h(X)\right\}\\
    &= \frac{1}{\theta}\mathbb{E}\left\{X(h(X) - h(X-1))\right\}
\end{align*}

\item Suppose $X\sim \mbox{Binom}(n,\theta)$, and let $h(X)$ be a function of $X$. Then we can show that 
\[\frac{\partial}{\partial \theta^2}\{\theta^2 \mathbb{E}[h(X)] \} = \frac{1}{\theta} \mathbb{E}\left[X\left\{(X+1)h(X) -2Xh(X-1) + (X-1)h(X-2) \right\}\right]\]

\textit{Proof}:
\begin{align*}
 \frac{\partial^2}{\partial \theta^2}\{\theta \mathbb{E}[h(X)] \} &= \frac{\partial}{\partial\theta}[\mathbb{E}[h(X)] +  \mathbb{E}[X\{h(X) -h(X-1)\}]  \\
 &= \frac{\partial}{\partial \theta} \mathbb{E}[(X+1)h(X) - Xh(X-1)]\\
&= \frac{1}{\theta}\mathbb{E}[X\{(X+1)h(X) - 2Xh(X-1) + (X-1)h(X-2)\}]
\end{align*}
\end{enumerate}

Consider the function $g:[0,\infty) \to \mathbb{R}$ and let $g(x) = 0$ for $x\in (-\infty, 0)$. Let $X = X(\theta)$ for simplicity of the notations. Using the results from the foregoing lemmas, we have

\begin{align*}
\frac{\partial^2 }{\partial \theta^2 }\mathbb{E}[g(X)] &=  \frac{\partial}{\partial \theta}\left(2\theta \mathbb{E}[g(X)] + \theta \mathbb{E}[X\{g(X) - g(X-1)\}]  \right) \\
&= 2\mathbb{E}[g(X)] + 2\mathbb{E}[X\{g(X)-g(X-1)\}] + \mathbb{E}[X\{g(X) - g(X-1)\}] \\
&+ \theta \frac{\partial}{\partial \theta} \mathbb{E}[X\{g(X) - g(X-1)\}]\\
&= 2\mathbb{E}[g(X)] + 2\mathbb{E}[X\{g(X)-g(X-1)\}] + \mathbb{E}[X\{g(X) - g(X-1)\}]\\
&+ \mathbb{E}\left[ X\{ X(g(X) - g(X-1))  - (X-1)(g(X-1) - g(X-2)) \}   \right]\\
&= \mathbb{E}[(X^2+3X +2) g(X) - (2X^2 + 2X)g(X-1) + (X^2-X)g(X-2)]\\
&= \mathbb{E}[\Tilde{g}(X) - 2\Tilde{g}(X-1) + \Tilde{g}(X-2)]
\end{align*}
where $\Tilde{g}(x) = (x+2)(x+1)g(x)$ for $x\in \mathbb{R}$.

Returning to the definition of $G(\theta)$, since $\alpha_{A_j} = \sum_{i\in A_j} \alpha_i$ which only depend on $|A_j|$ but not $\theta_{A_j}$, we write $\Tilde{\alpha} = \alpha_{A_j}$ for notation simplicity. Suppose that
\[g(x) = -\log \left\{1 - \frac{1}{1 + x + \Tilde{\alpha}}\right\}\]
for all $x\in [0,\infty)$. 

Fixing $x\in [0,\infty)$, we have that

\[\Tilde{g}(x) = (x^2 + 3x + 2)\log(\frac{1 + x + \Tilde{\alpha}}{x + \Tilde{\alpha}})\]

\[\Tilde{g}'(x) = (2x + 3) \log\frac{ 1 + x+ \Tilde{\alpha}}{x + \Tilde{\alpha}} - \frac{(x^2 + 3x + 2)}{(1+ x + \Tilde{\alpha})(x + \Tilde{\alpha})} \]

\[\Tilde{g}''(x) = 2 \log \frac{1 + x + \Tilde{\alpha}}{x + \Tilde{\alpha}} - 2\times \frac{(2x+3)}{(1+x+\Tilde{\alpha})(x+\Tilde{\alpha})} + \frac{x^2+3x + 2} {(1+x+\Tilde{\alpha})^2(x+\Tilde{\alpha})} + \frac{(x^2 + 3x +2)}{(1+x+\Tilde{\alpha})(x+\Tilde{\alpha})^2}\]

Hence by the low-integer-order polylogarithms series property $\sum_{k=1}^{\infty} \frac{1}{k}z^k = -\log(1-z)$ we have that

\begin{align*}
(1+x + \Tilde{\alpha})\Tilde{g}''(x) &=  2\sum_{k=1}^{\infty}\frac{1}{k}\frac{1}{(1+x+\Tilde{\alpha})^{k-1}} - 2\times \frac{2x + 3}{x + \Tilde{\alpha}} + \frac{x^2 +3x +2}{(1+x+\Tilde{\alpha})(x+\Tilde{\alpha})} + \frac{(x^2 + 3x + 2)}{(x+\Tilde{\alpha})^2}\\
&= 2 + 2\sum_{k=2}^{\infty} \frac{1}{k}\frac{1}{(1+x+ \Tilde{\alpha})^{k-1}} - 4\times (1-\frac{\Tilde{\alpha} - 3/2}{x+\Tilde{\alpha}}) \\
&+ \left(1-\frac{\Tilde{\alpha} - 1}{1 + x +\Tilde{\alpha}} + 1 - \frac{\Tilde{\alpha}-2}{x + \Tilde{\alpha}}\right)\left(1-\frac{\Tilde{\alpha} -1}{x+\Tilde{\alpha}}\right)\\
&= 2\sum_{k=2}^{\infty} \frac{1}{k}\frac{1}{(1+x+\Tilde{\alpha})^{k-1}} + 4\times \frac{\Tilde{\alpha} - 3/2}{x+\Tilde{\alpha}}\\
&- \left(\frac{\Tilde{\alpha}-1}{1+x+\Tilde{\alpha}} + \frac{\Tilde{\alpha}-2}{x+\Tilde{\alpha}}\right) - 2\times \frac{\Tilde{\alpha}-1}{x + \Tilde{\alpha}} +  \left(\frac{\Tilde{\alpha}-1}{1+x+\Tilde{\alpha}} + \frac{\Tilde{\alpha}-2}{x+\Tilde{\alpha}}\right)\frac{\Tilde{\alpha}-1}{x+\Tilde{\alpha}}\\
&= 2\sum_{k=2}^{\infty} \frac{1}{k}\frac{1}{(x+1+\Tilde{\alpha})^{k-1}} - \frac{1}{x+\Tilde{\alpha}} + \frac{\Tilde{\alpha}-1}{(x+\Tilde{\alpha})(1+x+\Tilde{\alpha})} + \left(\frac{\Tilde{\alpha}-1}{1+x+\Tilde{\alpha}} + \frac{\Tilde{\alpha}-2}{x+\Tilde{\alpha}}\right)
\end{align*}

If $\Tilde{\alpha} \geq 2$, then 

\begin{align*}
 (1+x + \Tilde{\alpha})\Tilde{g}''(x) &\geq 2\times \frac{1}{2} \times \frac{1}{1+x+ \Tilde{\alpha}} - \frac{1}{x + \Tilde{\alpha}} + \frac{\Tilde{\alpha}-1}{(x+\Tilde{\alpha})(1+x+\Tilde{\alpha})} \\
 &= -\frac{1}{(x+\Tilde{\alpha})(1+x+\Tilde{\alpha})} + \frac{\Tilde{\alpha}-1}{(x+\Tilde{\alpha})(1+x+\Tilde{\alpha})}\geq 0
\end{align*}

Therefore, $\Tilde{g}(X) - 2\Tilde{g}(X-1) + \Tilde{g}(X-2) \geq 0$ when $X\geq 2$.

For $X = 1$, we have 

\begin{align*}
    \frac{1}{6}(\Tilde{g}(1) - 2\Tilde{g}(0) + \Tilde{g}(-1)) &= \log\left(1+ \frac{1}{1+ \Tilde{\alpha}}\right) - \frac{2}{3}\log\left(1+\frac{1}{\Tilde{\alpha}}\right)\\
    &\geq \log\left(1+\frac{1}{1+\Tilde{\alpha}}\right) - \log\left(1 + \frac{2/3}{\Tilde{\alpha}}\right) \geq 0
\end{align*}

For $X=0$, we have $\Tilde{g}(0) - 2\Tilde{g}(-1) + \Tilde{g}(-2)\geq 0$ trivially. Then we have shown that 

\[\frac{\partial^2}{\partial \theta^2}\theta^2\mathbb{E}[g(X)] = \mathbb{E}[\Tilde{g}(X) - 2\Tilde{g}(X-1) + \Tilde{g}(X-2) ]\geq 0\]

Hence when $\alpha_{A_j}\geq 2$, the function $G$ is convex. 

Assuming $\alpha_{A_j}\geq 2$, because $G$ is convex and that $\Delta_{\bm\theta}(N,1)$ has convex domain, the maximum value exits and can be attained. We would like to find the $\bm\theta^*$ such that the risk difference $\Delta_{\bm\theta}(N,1)$ is maximized.

Let $\bm\theta^*  = (\theta_{i}^*)_{i=0}^k \in \bm\Theta$, where for $i\in A_j = \{ j_1, \dots, j_{n_j}\}$, $\theta_{i: i\in A_j}^* = \frac{1}{(1+m)n_{j}}$ for all $j = 0, \dots, m$. Then $\sum_{i=0}^k \theta^*_{i:i\in A_j} = \theta_{A_j} = \frac{1}{m+1}$ for all $j = 0, 1,\dots, m$.

By Jensen's inequality for real convex function, we have that 
\[\phi(\frac{\sum a_i x_i}{\sum a_i}) \leq \frac{\sum a_i \phi(x_i)}{\sum a_i}\]
and as a particular case, if $a_i$ are equal, we have 
\[\phi(\frac{\sum x_i}{n})\leq \frac{\sum \phi(x_i)}{n}\]

Therefore, 
\begin{align*}
\Delta_{\bm\theta}(N,1) & = \log\left\{ 1 + \frac{1}{N +\alpha_S }\right\} - \sum_{j=0}^m G(\theta_{A_j})\\
&=  \log\left\{ 1 + \frac{1}{N +\alpha_S }\right\} - (1+m)\frac{1}{1+m} \sum_{j=0}^m G(\theta_{A_j})\\
&\leq  \log\left\{ 1 + \frac{1}{N +\alpha_S }\right\} - (1+m) G(\frac{1}{m+1}\sum_{j=1}^m \theta_{A_j})\\
&=  \log\left\{ 1 + \frac{1}{N +\alpha_S }\right\} - \sum_{i=0}^m G(\frac{1}{m+1})\\
&= \Delta_{\bm \theta^*}(N, 1)
\end{align*}

With the foregoing results, we attempt to obtain the condition for the dominance between $\hat {\bm \theta}$ and $\Tilde{\bm \theta}$:

\begin{align*}
     \underset{\bm\theta\in\bm\Theta}{\mbox{sup}}\Delta_{\bm\theta}(N, N') &=  \underset{\bm\theta\in\bm\Theta}{\mbox{sup}}\sum_{u=1}^{N'}\Delta_{\bm \theta}(N + u - 1, 1)\\
     &\leq \sum_{u=1}^{N'}  \underset{\bm\theta\in\bm\Theta}{\mbox{sup}}\Delta_{\bm \theta}(N + u - 1, 1) \\
     &= \sum_{u=1}^{N'} \Delta_{\bm \theta^*}(N + u - 1, 1)\\
     &= \Delta_{\bm \theta^*}(N, N')
\end{align*}

Then 
\begin{align*}
\Delta_{\bm \theta^*}(N, N')   &=\mathbb{E}_{\bm \theta^*}\left[\log \frac{N' + N + \alpha_S}{N +\alpha_S }   +\sum_{j=0}^m \frac{1}{m+1}\log\left\{\frac{x_{A_j} + \alpha_{A_j}}{y_{A_j} +x_{A_j} + \alpha_{A_j} }\right\}\right]\\
&= -\mathbb{E}_{\bm \theta^*}\left[\sum_{j=0}^m \frac{1}{m+1} \log\left\{\frac{y_{A_j} + x_{A_j} + \alpha_{A_j}}{x_{A_j} + \alpha_{A_j}}\right\} + \log \frac{N + \alpha_S}{N'+N + \alpha_S}\right]\\
&= -\mathbb{E}_{\bm \theta^*}\left[ \sum_{j=0}^m \frac{1}{m+1} \log\left\{\frac{y_{A_j} + x_{A_j} + \alpha_{A_j}}{x_{A_j} + \alpha_{A_j}}\times \frac{N + \alpha_S}{N + N' + \alpha_S}\right\}\right]\\
&= -\mathbb{E}_{\bm \theta^*}\left[\sum_{j=0}^m \frac{1}{m+1}\log\left\{\frac{y_{A_j}/N' + (x_{A_j} + \alpha_{A_j})/N'}{1 + (N + \alpha_S)/N'}\bigg/ \frac{x_{A_j} + \alpha_{A_j}}{N + \alpha_S}\right\}\right]\\
&\to -\mathbb{E}_{\bm \theta^*}\left[\sum_{j=0}^m \frac{1}{m+1}\log\left\{\frac{1}{m+1}\bigg/ \frac{x_{A_j} + \alpha_{A_j}}{N + \alpha_S}\right\}\right]
\end{align*}
as $N'\to \infty$ by law of large numbers. We can show that the right-hand side is negative because 
\[\mathbb{P}_{\theta^*}\left\{\left(\frac{x_{A_j} +\alpha_{A_j}}{N + \alpha_S}\right)_{j=0}^m \neq \left(\frac{1}{m+1}\right)_{j=0}^m\right\}>0\]

Therefore, 
\[\Delta_{\bm \theta^*}(N, N') \leq {\mbox{sup}}\Delta_{\bm\theta}(N, N') <0 \]
for sufficiently large $N'$.

\subsection{Proof of Theorem 3}

Assume $N' = 1$ without loss of generality. Then it is sufficient to show that
\[\Delta_{\bm \theta^*}(N, 1) = \log\left\{1 + \frac{1}{N + \alpha_S} \right\}- (m+1)G(\frac{1}{m+1})\]
is negative when $N$ is sufficiently large. 

Let $\theta_{A_j}^* = \frac{1}{1+m}$ and $X\sim \mbox{Bin}(N, \theta_{A_j}^*)$

\begin{align*}
    (1+m)G(\frac{1}{1+m}) &= \frac{1}{1+m}\mathbb{E}\left[-\log\left\{1 - \frac{1}{1+ X + \alpha_{A_j}}\right\}\right]\\
    &= \theta_{A_j}^*\sum_{k=1}^{\infty}\frac{1}{k}\mathbb{E}\left[\frac{1}{(1+X+\alpha_{A_j})^k}\right]
\end{align*}

By (3) of Cressie et al. (1981), which is

\[\mathbb{E}(X^{-n}) = \Gamma(n)^{-1}\int_{0}^{\infty}t^{n-1}M_X(-t)dt\]

We have that 
\begin{align*}
\frac{1}{k}\mathbb{E}\left[\frac{1}{(1+X+\alpha_{A_j})^k}\right] &= \frac{1}{k}\frac{1}{\Gamma(k)}\int_{0}^{\infty} t^{k-1} \mathbb{E}[e^{-(1+X+\alpha_{A_j})t}]dt\\
&= \frac{1}{k!} \int_{0}^{\infty} t^{k-1} e^{-(1+\alpha_{A_j})t}\mathbb{E}[e^{-Xt}]dt
\end{align*}
for $k = 1, 2, 3,\dots$. Then 

\begin{align*}
    (m+1)G(\frac{1}{m+1}) &= \theta_{A_j}^*\int_{0}^{\infty}\sum_{k=1}^{\infty}\frac{t^{k-1}}{k!} e^{-(1+\alpha_{A_j})t} \mathbb{E}[e^{-Xt}]dt\\
    &=\theta_{A_j}^*\int_{0}^{\infty} \frac{e^t-1}{t}e^{-(1+\alpha_{A_j})t}\mathbb{E}[e^{-Xt}]dt
\end{align*}

Note that $\mathbb{E}[e^{-Xt}] = (1-\theta_{A_j}^* + \theta_{A_j}^*e^{-t})^N$ for all $t\in (0,\infty)$ by the moment generating function of binomial distribution. Then we have 

\begin{align*}
  (m+1)G(\frac{1}{m+1}) &=   \theta_{A_j}^*\int_{0}^{\infty} \frac{e^t-1}{t}e^{-(1+\alpha_{A_j})t}(1-\theta_{A_j}^* + \theta_{A_j}^*e^{-t})^Ndt
\end{align*}

Also,

\begin{align*}
\log\left\{1 + \frac{1}{N + \alpha_S} \right\}  &= -\log\left\{1-\frac{1}{1 + N + \alpha_S}\right\}=\sum_{k=1}^{\infty} \frac{1}{k}\frac{1}{(N + 1 + \alpha_S)^k}\\
&= \sum_{k=1}^{\infty}\frac{1}{k!}\int_{0}^{\infty} t^{k-1}e^{-(N + 1 + \alpha_S)t}dt\\
&= \int_{0}^{\infty} \frac{e^t - 1}{t}e^{-(N+1)t}e^{-\alpha_St}
\end{align*}

Therefore,

\begin{align*}
\Delta_{\bm \theta^*}(N, 1) &=  \int_{0}^{\infty} \frac{e^t - 1}{t}e^{-(N+1)t}e^{-\alpha_St} -\theta_{A_j}^*\int_{0}^{\infty} \frac{e^t-1}{t}e^{-(1+\alpha_{A_j})t}(1-\theta_{A_j}^* + \theta_{A_j}^*e^{-t})^Ndt
\end{align*}

Note that 

\begin{align*}
0 &\leq \left(\frac{e^t - 1}{t} \right)= \sum_{k=1}^{\infty}\frac{t^{k-1}}{k!} \leq \sum_{k=1}^{\infty} \frac{t^{k-1}}{(k-1)!} = e^t\\
0&\leq \left( \frac{e^t - 1}{t}\right)' = \sum_{k=2}^{\infty}\frac{(k-1)t^{k-2}}{k!} \leq \sum_{k=2}^{\infty}\frac{t^{k-2}}{(k-2)!} = e^{t}\\
0&\leq  \left( \frac{e^t - 1}{t}\right)'' = \sum_{k=3}^{\infty}\frac{(k-1)(k-2)t^{k-3}}{k!} \leq \sum_{k=3}^{\infty}\frac{t^{k-3}}{(k-3)!} = e^t
\end{align*}

and 
\begin{align*}
    \left(\frac{e^t - 1}{t}\right) &\leq \frac{e^t}{t} \\
      \left(\frac{e^t - 1}{t}\right)' &= \frac{e^t}{t} - \frac{e^t-1}{t^2} \leq \frac{e^t}{t}\\
        \left(\frac{e^t - 1}{t}\right)'' &= \frac{e^t}{t}-2\frac{e^t}{t^2} + 2\frac{e^t-1}{t^3} \leq \frac{e^{t}}{t} + 2\frac{e^t}{t^3}
\end{align*}

By integration by parts, we have 

\begin{align*}
  (N+1)\Delta_{\bm \theta^*}(N, 1) & = \left[-\frac{e^t-1}{t}e^{-(N+1)t}e^{-\alpha_St}\right]_{0}^\infty  - \left[- \frac{e^t-1}{t} e^{-\alpha_{A_j}t}(1 + \theta_{A_j}^* + \theta_{A_j}^* e^{-t})^{N+1}\right]_0^{\infty}\\
  &+ \int_{0}^\infty \left\{ \left(\frac{e^t - 1}{t}\right)' - \left(\frac{e^t-1}{t}\right)\alpha_S\right\} e^{-(N+1)t}e^{-\alpha_S t}dt\\
  &- \int_{0}^\infty \left\{  \left(\frac{e^t - 1}{t}\right)'  -\left(\frac{e^t-1}{t}\right)\alpha_{A_j} \right\} e^{-\alpha_{A_j}t}(1-\theta_{A_j}^* + \theta_{A_j}^*e^{-t})^{N+1}dt\\
  &= \int_0^\infty h_1(t) e^{-(N+1)t}e^{-\alpha_S t}dt - \int_0^\infty h_2(t)  e^{-\alpha_{A_j}t}(1-\theta_{A_j}^* + \theta_{A_j}^*e^{-t})^{N+1}dt
\end{align*}

where $h_1(t) =  \left(\frac{e^t - 1}{t}\right)' - \left(\frac{e^t-1}{t}\right)\alpha_S$ and $h_2(t) =  \left(\frac{e^t - 1}{t}\right)'  -\left(\frac{e^t-1}{t}\right)\alpha_{A_j} $ for $t\in (0,\infty)$.

By integration by parts, we have

\begin{align*}
     (N+1)^2\Delta_{\bm \theta^*}(N, 1) & = \left[h_1(t) e^{-(N+1)t}e^{-\alpha_S t}\right]_0^\infty \\
     &- \frac{N+1}{N+2}\left[\frac{1}{\theta_{A_j}^*}h_2(t)e^{-(\alpha_{A_j}-1)t}(1-\theta_{A_j}^* + \theta_{A_j}^*e^{-t})^{N+2} \right]_0^\infty\\
     &+ \int_0^\infty \left\{h_1'(t) - h_1(t) \alpha_S\right\}e^{-(N+1)t}e^{-\alpha_St}dt\\
     &- \frac{N+1}{N+2}\int_0^\infty \frac{1}{\theta_{A_j}^*} \left[h_2'(t)-h_2(t)(\alpha_{A_j}-1)\right]e^{-(\alpha_{A_j} - 1)t}(1-\theta_{A_j}^* + \theta_{A_j}^* e^{-t})^{N+2}dt\\
     &= h_1(0) - \frac{N+1}{N+2}\frac{1}{\theta_{A_j}^*}h_2(0) + \int_{0}^\infty \left[h_1'(t) - h_1(t)\alpha_S\right]e^{-(N+1)t}e^{-\alpha_St}dt\\
     &- \frac{N+1}{N+2}\int_0^\infty \frac{1}{\theta_{A_j}^*}\left[h_2'(t) - h_2(t)(\alpha_{A_j}-1)\right]e^{-(\alpha_{A_j}-1)t}(1-\theta_{A_j}^* + \theta_{A_j}^* e^{-t})^{N+2}dt
\end{align*}

Consider the following
\begin{align*}
\bigg| \left[h_2'(t) - h_2(t)(\alpha_{A_j}-1)\right]e^{-(\alpha_{A_j}-1)t} \bigg| &\leq \left[e^t + e^t\alpha_{A_j}(\alpha_{A_j}-1)\right]e^{-(\alpha_{A_j}-1)t} \\
& = [1+ \alpha_{A_j}(\alpha_{A_j}-1)]e^{-(\alpha_{A_j}-2)t}
\end{align*}

By assumption we have $\alpha_{A_j}\geq 2$.

For $\alpha_{A_j} > 2$, when $t\in (0,1]$, we have 

\[\bigg| \left[h_2'(t) - h_2(t)(\alpha_{A_j}-1)\right]e^{-(\alpha_{A_j}-1)t} \bigg| \leq \bigg| h_2'(t) - h_2(t)\bigg| = 6e^t\]

and when $t \in (1,\infty)$ we have 

\[\bigg| \left[h_2'(t) - h_2(t)(\alpha_{A_j}-1)\right]e^{-(\alpha_{A_j}-1)t} \bigg| \leq \frac{3}{t^2} + \frac{7}{t}e^{-t}\]

For $\alpha_{A_j} = 2$, by the dominated convergence theorem, as $N\to \infty$,

\[(N+1)^2 \Delta_{\bm\theta^*}(N,1) \to h_1(0) - \frac{1}{\theta_{A_j}^*}h_2(0) = \frac{1}{2}(1 - \frac{1}{\theta_{A_j}^*}) <0\]

The maximum risk difference is 

\begin{align*}
\Delta_{\bm \theta^*}(N, 1) &= \log\left\{1 + \frac{1}{N + \alpha_S} \right\}- (m+1)G(\frac{1}{m+1})\\
&= \log\left\{1 + \frac{1}{N + \alpha_S} \right\} - \frac{1}{m+1}\mathbb{E}\left[-\log \left\{1-\frac{1}{1+X +\alpha_{A_j}}\right\}\right]
\end{align*}

\subsection{Additional Results}

\begin{figure}[ht]
\centering
\caption*{True 1996-2005 MCHSS Data}
\includegraphics[width = \linewidth]{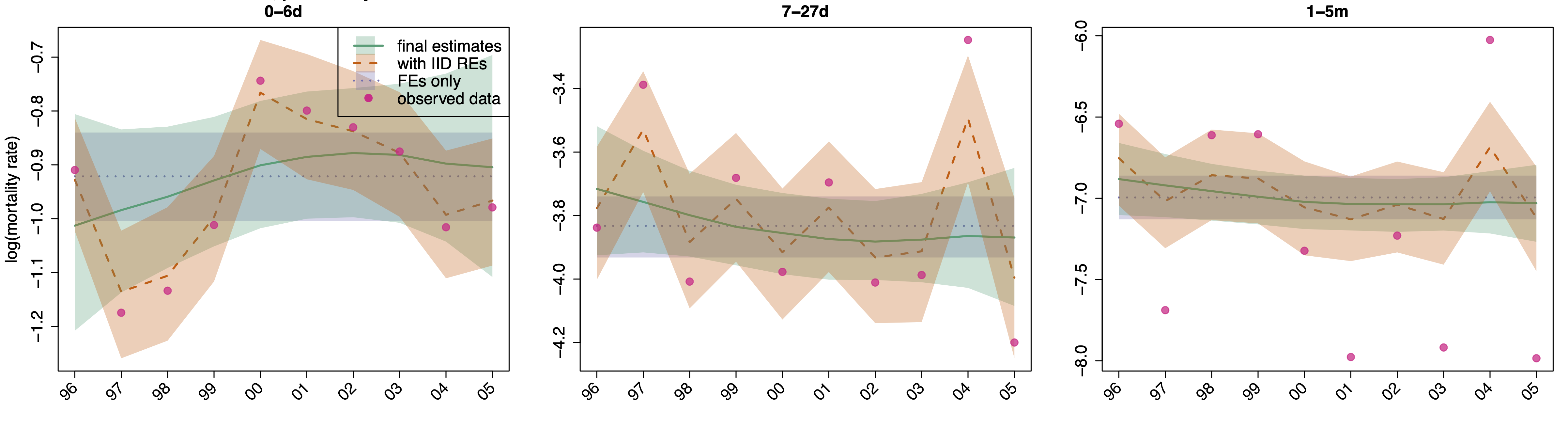}

\caption*{Predicted 1996-2005 MCHSS Data}

\includegraphics[width = \linewidth]{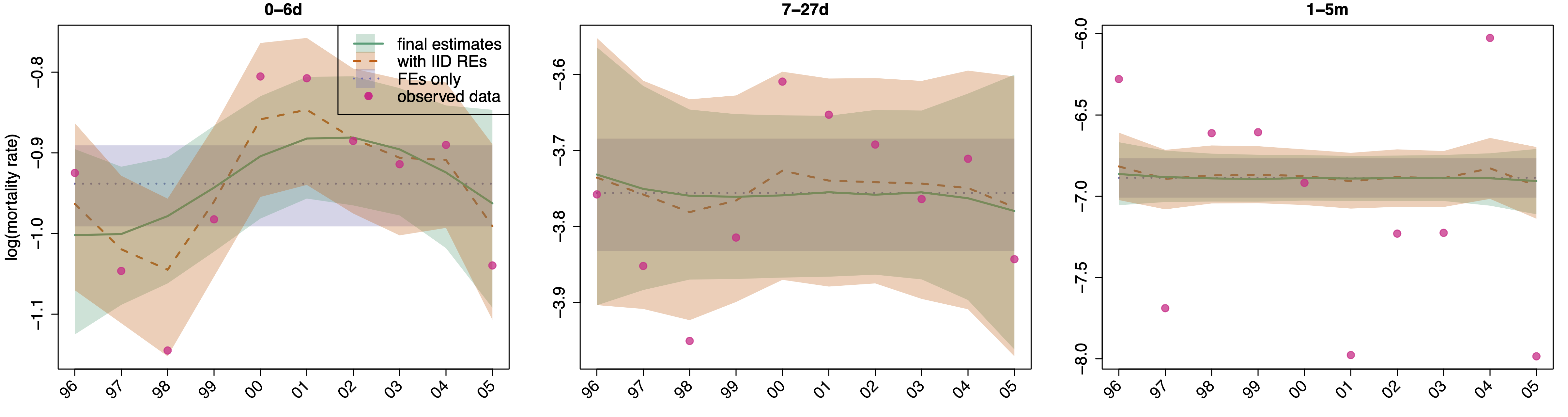}
\caption{Selected results from the MCHSS data showing empirical data, estimated posterior medians, and posterior $80\%$ intervals for log mortality rates of prematurity in the west rural region. Combinations with no deaths are represented by an open square. }
\label{fig:logmx-compare-west-rural-add}
\end{figure}

\begin{figure}[ht]
\centering
\caption*{True 1996-2005 MCHSS Data}
\includegraphics[width = \linewidth]{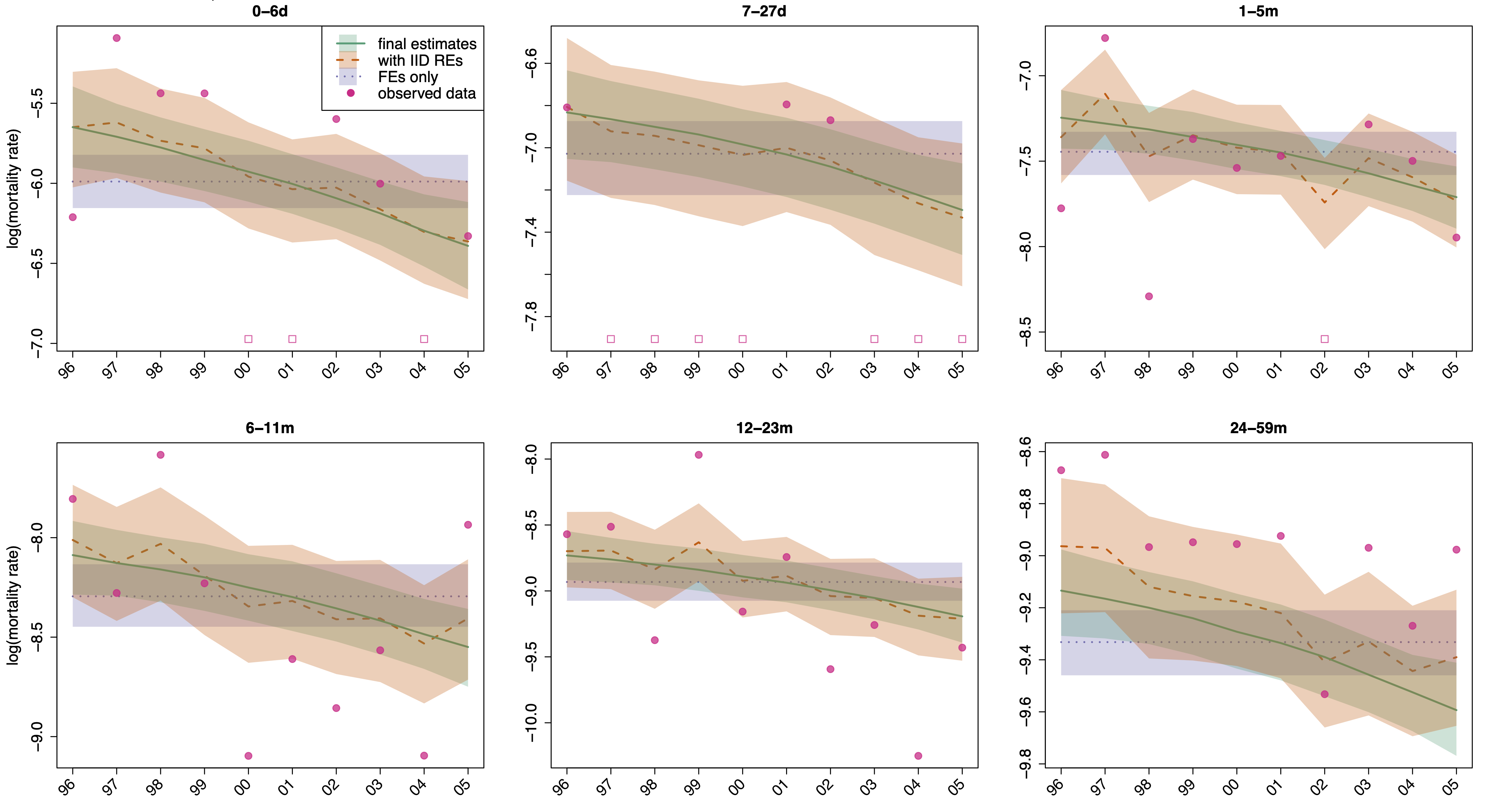}

\caption*{Predicted 1996-2005 MCHSS Data}
\includegraphics[width = \linewidth]{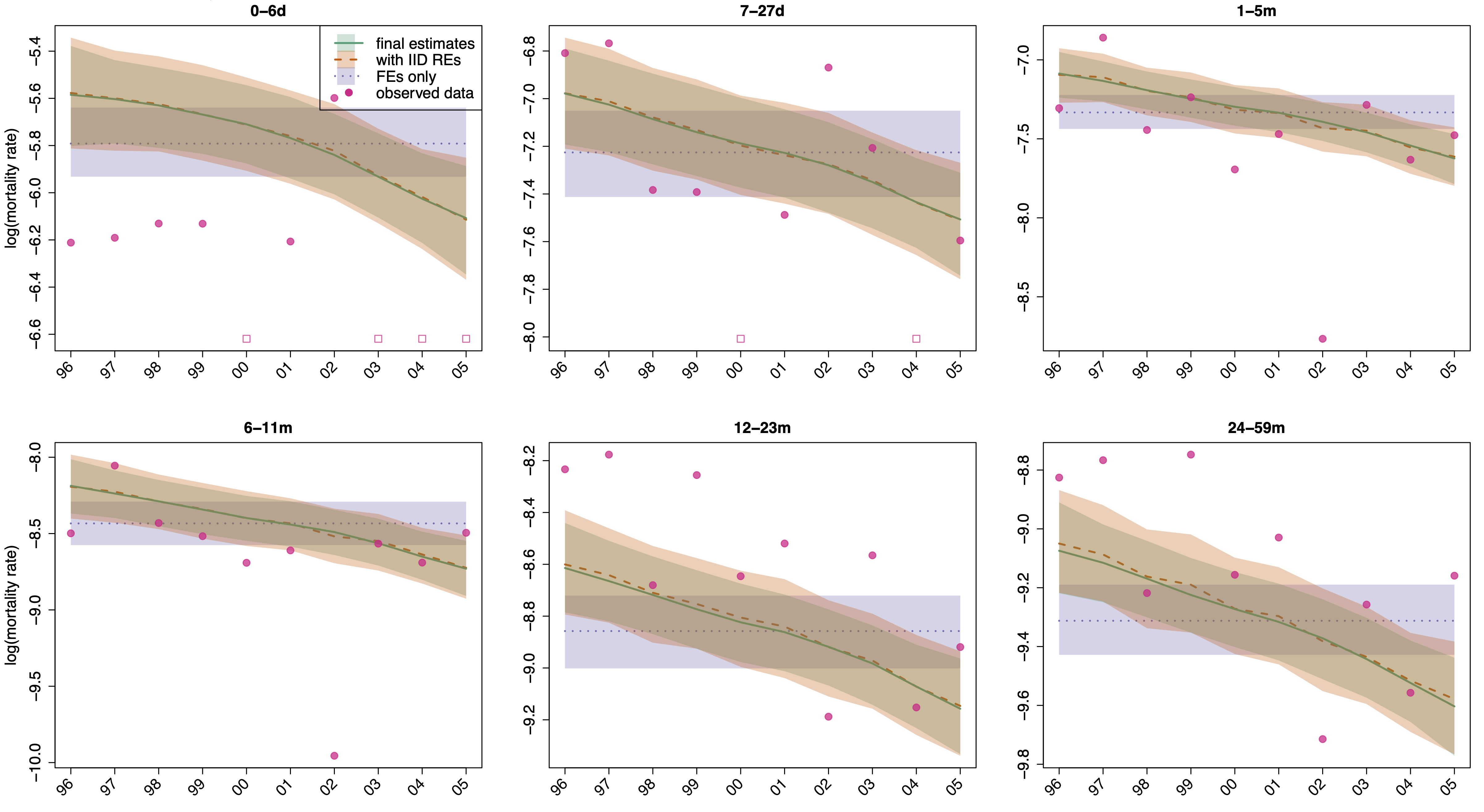}

\caption{Selected results from the MCHSS data showing empirical data, estimated posterior medians, and posterior $80\%$ intervals for log mortality rates of other non-communicable diseases in the east urban region. Combinations with no deaths are represented by an open square. }
\label{fig:logmx-compare-east-urban-add}
\end{figure}

\begin{figure}[ht]
\centering
\caption*{True 1996-2005 MCHSS Data}
\includegraphics[width=10cm,height=9cm]{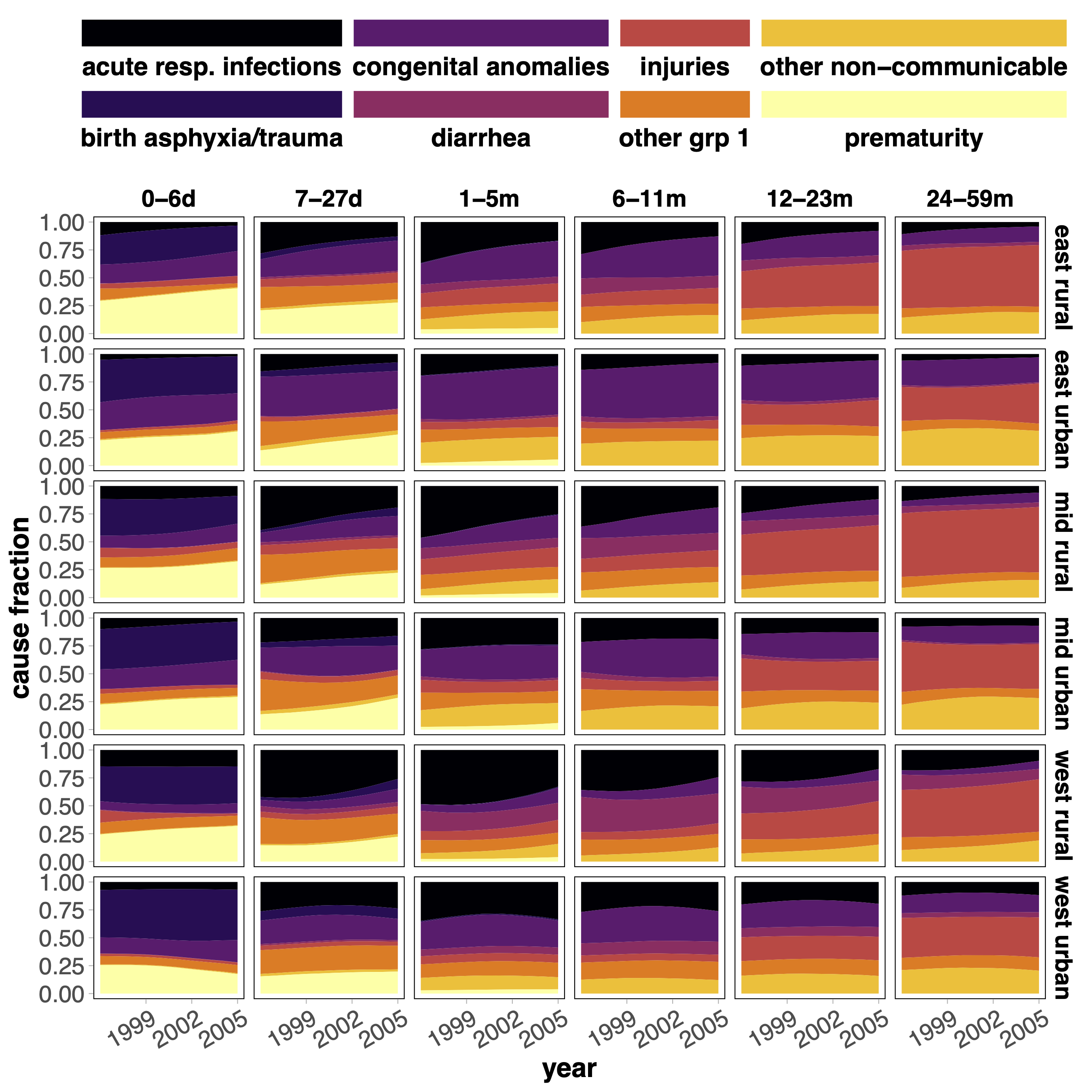}\\
\caption*{Predicted 1996-2005 MCHSS Data}
\includegraphics[width=10cm,height=8cm]{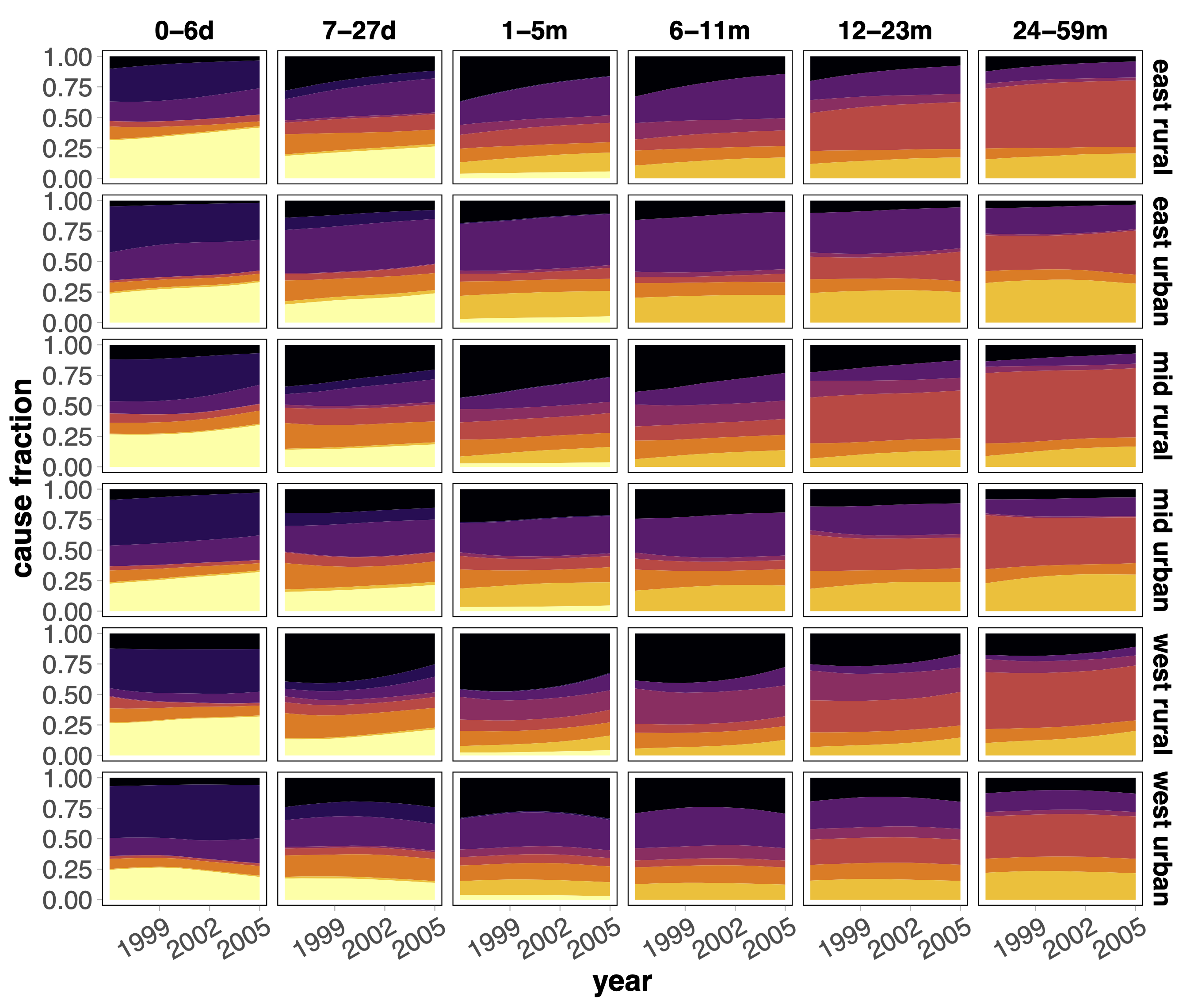}
\caption{Comparisons of estimated CSMFs between models based the true MCHSS data and the estimated MCHSS data using the proposed data reconciliation method. }
\label{fig:csmf-compare-full}
\end{figure}

\end{document}